\documentclass[10pt,a4paper,english,oneside]{article}
\usepackage[english]{babel}
\usepackage[utf8]{inputenc}
\usepackage[margin=1.5in]{geometry}
\usepackage{xcolor}
\usepackage{amsmath,amsfonts}
\usepackage{graphicx}
\usepackage{subcaption}
\usepackage{comment}
\usepackage{booktabs}
\usepackage{authblk}
\usepackage{bm}
\usepackage{amsthm}
\usepackage{tabularx}
\usepackage{makecell}
\usepackage{csquotes} 

\theoremstyle{definition}
\newtheorem{definition}{Definition}[section]

\usepackage{biblatex}
\usepackage[breaklinks]{hyperref}
\hypersetup{
    colorlinks,
    linkcolor={red!80!black},
    citecolor={green!70!black},
    urlcolor={blue!80!black}
}

\title{From chambers to echo chambers: Quantifying polarization with a second-neighbor approach applied to Twitter's climate discussion}

\author[1,2,$\dagger$]{Blas Kolic}
\author[3,4,5, $\diamond$]{Fabián Aguirre-López}
\author[6]{Sergio Hernández-Williams}
\author[6]{Guillermo Garduño-Hernández}

\affil[1]{{Institute for New Economic Thinking at the Oxford Martin School, University of Oxford, OX2 6ED Oxford, UK}}
\affil[2]{{uc3m-Santander Big Data Institute, Universidad Carlos III de Madrid, 28903 Madrid, Spain}} 
\affil[3]{{Université Paris-Saclay, CNRS, LPTMS, 91405, Orsay, France, 91405 Orsay, France}}
\affil[4]{Chair of Econophysics and Complex Systems, École polytechnique, 91128 Palaiseau, France}
\affil[5]{LadHyX UMR CNRS 7646, École polytechnique, 91128 Palaiseau, France}
\affil[6]{Sinnia, Mexico City, Mexico}
\affil[$\dagger$]{corresponding: \url{blas.kolic@uc3m.es}}
\affil[$\diamond$]{\url{fabian.aguirre-lopez@ladhyx.polytechnique.fr}}

\begin{document}

\newcommand{\red}[1]{\textcolor{red}{ #1}}
\newcommand{\blue}[1]{\textcolor{blue}{ #1 }}
\newcommand{\green}[1]{\textcolor{green}{ #1 }}
\newcommand{\bW}{\mathbf{W}}
\newcommand{\topusers}{\mathcal{I}}
\newcommand{\rank}{\textrm{rank}}
\newcommand{\bw}{\mathbf{w}}
\newcommand{\indicator}[1]{\mathbb{I}\left[ #1 \right]}
\newcommand{\calA}{\mathcal{A}}
\newcommand{\abs}[1]{\left| #1 \right|}
\newcommand{\bU}{\mathbf{U}}
\newcommand{\calC}{\mathcal{C}}
\newcommand{\bA}{\mathbf{A}}
\newcommand{\s}[1]{\left[ #1 \right]}
\newcommand{\p}[1]{\left( #1 \right)}
\newcommand{\expected}[1]{\left\langle #1 \right\rangle}
\newcommand{\chamber}{\mathcal{C}}
\newcommand{\Prob}{\textrm{Prob}}
\newcommand{\bk}{\mathbf{k}}
\newcommand{\audience}{\mathcal{A}}
\newcommand{\rme}{\mathrm{e}}
\newcommand{\rmi}{\mathrm{i}}
\newcommand{\rmd}{\textrm{d}}
\newcommand{\Var}{\textrm{Var}}
\newcommand{\users}{\mathcal{U}}
\newcommand{\calP}{\mathcal{P}}
\newcommand{\kin}{k^{\textrm{in}}}
\newcommand{\kout}{k^{\textrm{out}}}

\maketitle

\begin{abstract}
    Social media platforms often foster environments where users primarily engage with content that aligns with their existing beliefs, thereby reinforcing their views and limiting exposure to opposing viewpoints. In this paper, we analyze X (formerly Twitter) discussions on climate change throughout 2019, using an unsupervised method centered on \textit{chambers} --second-order information sources-- to uncover ideological patterns at scale. Beyond direct connections, chambers capture shared sources of influence, revealing polarization dynamics efficiently and effectively. 
    Analyzing retweet patterns, we identify echo chambers of climate believers and skeptics, revealing strong chamber overlap within ideological groups and minimal overlap between them, resulting in a robust bimodal structure that characterizes polarization. Our method enables us to infer the stance of high-impact users based on their audience’s chamber alignment, allowing for the classification of over half the retweeting population with minimal cross-group interaction, in what we term \textit{augmented echo chamber classification}. We benchmark our approach against manual labeling and a state-of-the-art latent ideology model, finding comparable performance but with nearly four times greater coverage. 
    Moreover, we find that echo chamber structures remain stable over time, even as their members change significantly, suggesting that these structures are a persistent and emergent property of the system. Notably, polarization decreases and climate skepticism rises during the \#FridaysForFuture strikes in September 2019. This chamber-based analysis offers valuable insights into the persistence and fluidity of ideological polarization on social media.
\end{abstract}

\section{Introduction}

Social media is crucial for information consumption and public opinion formation \cite{cinelli2021echo, gaumont2018reconstruction, garimella2018quantifying, bovet2018validation}. It leverages communication channels between one-to-many and many-to-many in a decentralized way by enabling its users to choose whom to follow and interact with, thus democratizing information access and spreading \cite{benkler2008wealth, ceron2016first}. However, the decentralized nature of social media encourages users to consume information mostly aligned with their beliefs and interact with like-minded individuals and communication channels. Authors have studied these behaviors under the sociological frameworks of confirmation bias \cite{wason1960failure} and homophily \cite{mcpherson2001birds}. Additionally, social media platforms are designed to maximize engagement time, with recommendation algorithms trained to show content similar to what users typically consume \cite{eirinaki2018recommender}. To a certain extent, the aforementioned mechanisms explain an empirical observation about controversial topics in social media: public discourse is polarized. Authors have studied polarization for several relevant topics, ranging from climate change \cite{cody2015climate, williams2015network, chen2021polarization, falkenberg2022growing}, to politics \cite{gaumont2018reconstruction,bovet2018validation,roth2021quoting}, to Covid-19 \cite{dyer2020public, jiang2021social,garcia2021extracting}.

According to Caves' \textit{Encyclopedia of the City} \cite{caves2004encyclopedia}, social polarization is the segregation within a society that emerges from several socio-economic factors that differentiate social groups. In the context of public discourse in social networks, \textit{echo chambers} reflect a parallel mechanism to social polarization. An echo chamber emerges, according to Bruns et al. \cite{bruns2017echo}, when a group of actors chooses to preferentially connect, with the exclusion of outsiders. However, over which preferences would such actors choose to connect? Garimella et al. \cite{garimella2018quantifying} state that opinions or beliefs stay inside communities created by like-minded people who reinforce and endorse each other's opinions. Therefore, agents prefer to connect to other actors with similar opinions and exclude those with contrasting ones, suggesting that \textit{homophily} is a driving force in the creation of echo chambers \cite{gillani2018me, mcpherson2001birds}. An actor inside an echo chamber will mostly receive information consistent with their beliefs, so they will reject any new information that does not align with those beliefs. In other words, actors in an echo chamber experience a \textit{confirmation bias}. This bias creates a positive feedback mechanism that reinforces the echo chambers in the light of new information \cite{sikder2020minimalistic}. 

Echo chambers inhibit communication across groups with different ideologies. Thus, even when a scientific consensus about a topic is reached, it will hardly permeate all the echo chambers within a social group, as is the case with climate change. According to the latest Intergovernmental Panel on Climate Change (IPCC) report \cite{ipcc2022summary}, the cumulative scientific evidence is unequivocal: Climate change threatens human well-being and planetary health.
Despite this, a significant fraction of the population in many countries express some form of doubt or skepticism about whether trends in climate change are anthropogenic, will bring harmful consequences to society, or both \cite{poortinga2019climate, kulin2021nationalist}. Such skepticism may be linked with factors such as the spread of misinformation online \cite{treen2020online}, nationalist and individualist identities \cite{kulin2021nationalist}, or, particularly in the United States, conservatism and conspiratorial beliefs \cite{hornsey2018relationships}. The issue of climate change is not only polarizing but also has deep social, economic, and ecological impacts. Shifts in public opinion may lead to significant behavioral and political changes, which could face strong opposition from specific population sectors.

Several authors have studied the climate opinion landscape in social media, finding, in most cases, that a group of \textit{climate believers}, i.e., actors that support the science about climate change, and a group of \textit{climate skeptics} coexist and segregate into online \textit{echo chambers}  \cite{williams2015network,tyagi2020polarizing,jang2015polarized,chen2021polarization}. Online social media has been a crucial driver of pro-climate movements, such as the ``Fridays for Future'' movement started by Greta Thunberg \cite{taylor2019climate, laville2019across}. Using Twitter (now X) data, Williams et al. \cite{williams2015network} constructed several interaction networks from relevant climate-related hashtags and manually labeled the most active users as believers, skeptics, or neutral. They measured high levels of homophily in the followers and retweet networks, suggesting that the conversation is polarized. With a similar approach, Jang and Hart \cite{jang2015polarized} found a polarized Twitter conversation on climate change in the United States between Republicans and Democrats. They claim that "climate change" and "global warming" are meaningful query filters for studying the climate conversation, with the latter being more common among climate skeptics. Chen et al. \cite{chen2021polarization} studied Finland's Twitter conversations about climate change by creating interaction networks from retweets. They found that the climate conversation was subject to partisan sorting and aligned with the universalist-communitarian dimension of European politics. Xia et al. \cite{xia2021spread} found that viral climate topics around the 2019 Peace Nobel Prize spread among different groups, enhancing in-group connections and repelling out-group engagement, suggesting the presence of echo chambers.

In this paper, we study the structure of the X (formerly Twitter) climate change conversation in 2019, when the "Fridays for Future" and "Extinction Rebellion" social movements flourished. We introduce unsupervised methods to 1) identify the leading users of the conversation throughout the year, 2) measure the ideological similarity between leading users by examining the many-to-many communication channels--or chambers--of the audience of a leading user, 3) determine the ideology of the leading users using their ideological similarities, and 4) present an operational definition of echo chamber based solely on the structure of the X interaction networks for which we classify more than half of the total retweeting population. We construct these methods assuming that retweeting is a good proxy for endorsement \cite{gaumont2018reconstruction, barbera2015birds}, and that information flows in the opposite direction of retweets. Moreover, we acknowledge that X conversations are highly heterogeneous, in that a tiny number of users account for most of the retweets produced by the population \cite{glenski2018propagation}. These methods build on prior work that defines echo chambers operationally \cite{cinelli2021echo} and models the ideological positions of active users on X \cite{barbera2015birds}. They also draw on vertex similarity measures that account for higher-order interactions, as many-to-many communication resembles second-order neighborhoods \cite{leicht2006vertex, malliaros2013clustering}.

We organize the remainder of this paper as follows: In section \ref{sec:leading_and_chambers}, we identify the \textit{leading users} of the climate conversation, introduce the \textit{chamber} associated with a leading user, and identify two polarized ideological groups based on the similarity of the chambers of the leading users using unsupervised methods. In section \ref{sec:echo_chambers}, we introduce an operational definition of an echo chamber and classify most of the users in the retweet network as either climate believers or climate skeptics. We inspect the properties of echo chambers and discuss communication within and across these groups. Finally, in section \ref{sec:conclusions}, we conclude and suggest future research directions. All code used in this paper is written in \textit{Python} and is openly available at \url{https://github.com/blas-ko/Twitter_chambers}. The repository includes scripts to fully reproduce all analyses and figures presented in the paper.

\section{The chambers of the X (previously Twitter) conversation} \label{sec:leading_and_chambers}

X users often consume information from various sources. They typically follow, audit, and interact with broadcasters that represent the one-to-many communication channels of the X-sphere and with other lower-impact users that represent the many-to-many communication channels \cite{liang2019did, goel2016structural, cha2010measuring}. In this paper, we introduce quantitative methods that leverage the impact of broadcasters, referred to as \textit{leading users}, and the set of lower-impact users, referred to as \textit{chambers}, within a given audience. Later, we introduce a quantitative definition of \textit{echo chamber} based solely on the structure of the X interaction networks. We focus on the climate change X conversation in 2019, where, based on past research \cite{williams2015network,tyagi2020polarizing,jang2015polarized,chen2021polarization}, we expect to find echo chambers of climate believers and climate skeptics. These methods characterize the dynamics and structure of the conversation in an \emph{unsupervised} way. Thus, we avoid missing certain features that could be ignored by our biases. Most importantly, these methods apply to other data where we do not have clear expectations of its structure. 

In this section, we define a \textit{leading user} and its associated \textit{chamber}--a second-order neighborhood-- and derive methods that identify polarized ideological groups in an unsupervised manner. We validate \textit{a posteriori} that such ideological groups correspond to those who believe in anthropogenic climate change and those who are skeptical about it.

\subsection{Data}
Alongside the company \textit{Sinnia}\footnote{\url{http://www.sinnia.com/en/}}, we create a dataset with $41.8$ million climate-related tweets from $8.7$ million users spanning 1$^{st}$ March to 1$^{st}$ December 2019, totalling $39$ weeks. We collect \textit{all} the tweets that satisfy any of the following queries: ``\textit{climate change}'', ``\textit{global warming}'', ``\textit{climatechange}'' and ``\textit{globalwarming}''. We choose these queries following the results from prior studies \cite{veltri2017climate, williams2015network, jang2015polarized}. Of the total volume of tweets, $73 \%$ corresponds to \emph{retweets}. We, therefore, focus on the retweets only, assuming that they provide significant information about what users are looking at or interested in. 

\subsection{High-impact and leading users \label{sec:leading-users}}

Discussions in X are highly heterogeneous \cite{sadri2018analysis}. This means that just a handful of users generate a huge fraction of the information consumed and spread through X interaction networks. However, these networks often exhibit community-like structures where the density of interactions is higher for users with similar features. Throughout this paper, we assume that retweets are a good proxy for endorsements \cite{barbera2015birds, gaumont2018reconstruction, chen2021polarization}, allowing us to study the retweet interaction network as a social network of endorsements. Thus, we study the dynamics of the retweet network, $\bW^t$, which we construct following Beguerisse et al. \cite{beguerisse2017and} as
\begin{align}
   W_{ij}^t =& \textrm{ \# of retweets of user $i$ originally posted by $j$ in a  given week $t$} ,\label{eq:def-Wij}\\
   w_i^t =&\sum_{j} W_{ji}^t = \textrm{\# of retweets of tweets posted by user $i$ in week $t$} ,
\end{align}
where $i$ and $j$ denote users from the complete set of retweeting users, $\users$, and $t$ denotes time, where we choose the \emph{week} as our temporal resolution. 

In this context, we refer as $w_i^t$ to the \textit{impact} of user $i$ during week $t$. From a network's perspective, the impact is the same as the weighted in-degree of a given node. Using the impact vector, $\bw^t = (w_i^t)$, we can quantify the inequality of the discussion in X by treating $w_i^t$ as a measure of user $i$'s wealth and the Gini index of its impact vector. The Gini index is $0$ if every user receives the same number of retweets and $1$ if a single user receives all the retweets.

The Gini index gives us a global picture of the impact heterogeneity, but we can also use $\bw^t$ to identify important users in the conversation. We define the set of $N$ \emph{high-impact users} in week $t$ as the $N$ most retweeted users during that week, i.e.,  
\begin{align}
    \topusers(t) = \{ i \in \users \mid \rank(w_i^t) \le N\}.
\end{align}
Importantly, $\topusers(t)$ is a dynamic set, as $rank(w_i^t)$ changes weekly for all users. This contrasts with previous work \cite{glenski2018propagation}, in which the set of high-impact users is fixed a priori.

The set $\topusers(t)$ can vary significantly during the weeks, so, to characterize the most important actors in the conversation, we introduce the \textit{persistence}, $\Delta_i$, as the number of weeks where $i$ is a high-impact user. Persistence allows us to identify relevant, high-impact users over longer time windows. We consider users with high impact and persistence as \emph{leading users} because they serve as one-to-many information channels relevant to focused audiences for extended periods \cite{liang2019did}. We thus define the set of $M$ \emph{leading users} as 
\begin{align}
    \topusers^\Delta (t) = \{ i \in \topusers(t)\mid \rank(\Delta_i)\le M\},
    \label{eq:persistent_leading_users}
\end{align}
that have an associated vector of impact,
\begin{align}
    \bw_{\topusers^\Delta}^t = (w_i^t)_{i\in\topusers^\Delta(t)}.
\end{align}

\begin{figure}[ht!]
    \centering
	\includegraphics[width=0.99\textwidth]{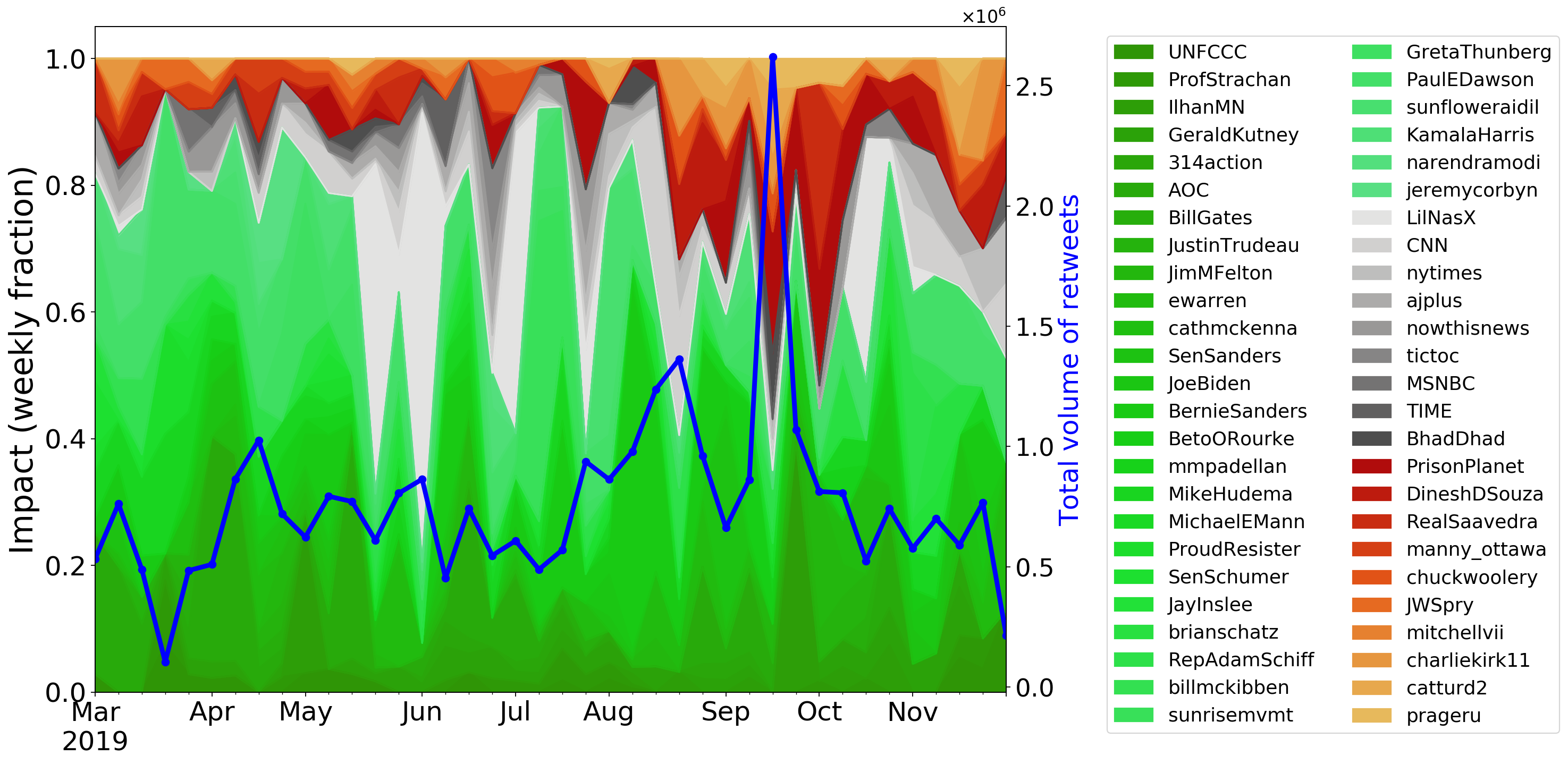}
	\caption{\textbf{Leading users popularity popularity dynamics} normalized to each week. We manually label green users as \textit{climate believers}, red users as \textit{climate skeptics}, and gray users as \textit{other}, such as news media and artists. In blue, we show the total volume of retweets for the given week, where we identify two significant peaks: one in the 3rd week of August and the other in the 3rd week of September 2019.}    
	\label{fig:impact_dynamics}    
\end{figure}

When we apply these definitions to our X dataset on climate change, we observe that almost $50\%$ of the weekly retweets are given to the top $N=50$ users. This is a massive inequality, as just a handful of tweets produce most interactions. We observe an average Gini index of $0.89$ ($\pm 0.02$)\footnote{Unless otherwise stated, quantities with uncertainty bars represent the standard deviation around the mean across weeks.}, which means that we should be able to understand a significant fraction of the X conversation just by looking at its high-impact users.

We take the $M = 50$ leading users from the $N=50$ high-impact weekly users. While the set of leading users, $\topusers^\Delta(t)$, is dynamic, in the analyses that follow, we focus on the complete set of leading users, $\topusers^\Delta = \cup_t \topusers^\Delta(t)$ where $\lvert \topusers^\Delta \rvert = 50$. As a first approach to understanding the set of leading users, we blue{\textit{manually assign}} each of them a category about climate change (see Appendix \ref{app:manual_labeling} for details on our manual labeling protocol). We identify three main ideological currents: \textit{climate believers} that support the scientific consensus on anthropogenic climate change, \textit{climate skeptics} that doubt that the trends in climate are anthropogenic or that such trends will bring harmful consequences for society, and \textit{news media channels and entertainers}. In Fig. \ref{fig:impact_dynamics}, we show the impact dynamics of the leading users \emph{per} week. While climate believers (green) generally dominate throughout the year, climate skeptics (red) surge noticeably in the second half, peaking in the third week of September 2019. Furthermore, Table \ref{tab:leading users_description} from Appendix \ref{app:leading users_description} lists the top $M=50$ leading users and their main characteristics.

\subsection{The chamber: quantifying leading users' similarity}

In the previous section, we presented a method for identifying the (small) set of \textit{leading users} who drive most of the X (previously Twitter) conversation on climate change. We manually labeled these users by determining their ideological stance based on information from their X bios, Wikipedia pages, and relevant news sources. While effective, manual labeling is time-consuming and requires outside information. Here, we introduce an unsupervised approach to uncover ideological (dis)similarities among leading users. This method analyzes the many-to-many communication channels within an audience, following interaction chains in a pattern similar to viral spread \cite{liang2019did}.

Each leading user has an associated \textit{audience} who endorses her, where the content the leading user posts flows in the direction of her audience, creating a one-to-many communication channel, similar to broadcasting \cite{goel2016structural}. Simultaneously, the audience consumes and endorses posts from other X users, forming a many-to-many communication channel, which we call the \textit{chamber} of such an audience. The audience and the chamber associated with a leading user are our primary subjects of study, and we introduce them in detail below.

\begin{definition}[Audience]
Given a user $i \in \users$, her \textit{audience} at week $t$, $\calA_i^t$, is the set of users that have retweeted $i$ during $t$:
\begin{equation}
    \calA_i^t = \{ j \in \users \mid W_{ji}^t >0\}.
    \label{eq:audience}
\end{equation}
\end{definition}

The set $\calA_i^t$ indicates who endorsed $i$ during week $t$. Thus, the ideologies of the members of $\calA_i^t$ are coherent with those of $i$'s. Additionally, the audience may interact with and endorse other lower-impact sources, thereby creating an information flow from these sources to the audience. We refer to this collection of sources as the \textit{chamber} of the audience associated with the leading user $i$.
\begin{definition}[Chamber]
    Given an audience, $\calA_i^t$, associated with a user $i \in \users$, the \emph{chamber} of $\calA_i^t$ during week $t$, $\calC^t_i$, is the set of users retweeted by the audience of $i$ excluding $i$:
\begin{equation}
    \calC^t_i = \{j \in \users \mid W_{kj}^t >0\textrm{ for } k\in\calA_i^t, \  k \neq i\}.
    \label{eq:chamber}
\end{equation}
\end{definition}

\begin{figure}[ht!]
    \centering
    \includegraphics[width=0.7\textwidth]{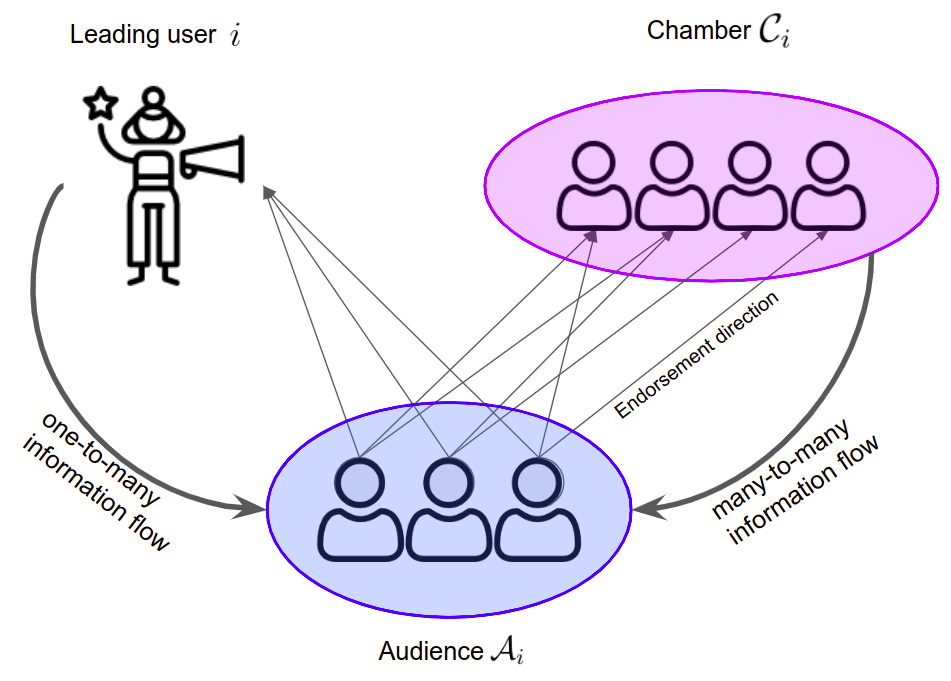}    
    \caption{\textbf{Schematic of the audience and the chamber of leading user $i \in \topusers^\Delta(t)$}. The audience, $\mathcal{A}_i^t$, is the set of users that retweet $i$ while the chamber, $\mathcal{C}_i^t$, is the set of users retweeted by every member of the audience, i.e., the \textit{sources of information} of the audience other than the leading user.}
    \label{fig:diagram_chamber}    
\end{figure}

The audience is a collection of information consumers, while the chamber is a collection of sources. See Fig. \ref{fig:diagram_chamber}, where we show a schematic representation of the audience and the chamber associated with a leading user. A leading user relates to its chamber through its audience, so we expect that the chamber members have an ideology similar to that of the leading user. We consider the (dynamic) audiences and chambers associated with the leading users, $\topusers^\Delta$, throughout the year. 

The chambers transmit information from the sources to the audiences. Thus, we can estimate the information flowing between two audiences associated with the leading users $i$ and $j$ by comparing the similarities between their chambers. Following our assumption that retweets indicate endorsement, this similarity serves as a proxy for the ideological distance between $i$ and $j$. Thus, we can compare the ideologies of two leading users by looking at their chamber overlaps, which we introduce below.
\begin{definition}[Chamber overlap]
Given two leading users $i$ and $j$ with respective chambers $\calC_i^t$ and $\calC_j^t$, the \textit{chamber overlap}, $q_{ij}^t$, during week $t$ is the Jaccard similarity between $\calC_i^t$ and $\calC_j^t$:
\begin{align}
\label{eq:jaccard}
    q_{ij}^t = \frac{\abs{\calC_i^t\cap\calC_j^t}}{\abs{\calC_i^t\cup\calC_j^t}} = \frac{ \textrm{\# of users in common of both chambers}}{\textrm{total \# of users of both chambers}}.
\end{align}
\end{definition}

The overlap between the chambers provides a proxy for the ideological similarity between $i$ and $j$ \textit{without} explicitly determining their ideological positions, meaning that $q_{ij}^t$ is a relative measure \textit{between} the two users. Intuitively, overlaps near zero indicate strong ideological separation, suggesting polarization due to minimal interaction between chambers. We quantify this intuition later in Section~\ref{sec:echo_chambers}. We could compare \textit{audience overlaps} using audiences $\calA_i^t$ instead of chambers in Eq. (\ref{eq:jaccard}). Still, we prefer to use the chambers for mathematical and conceptual arguments, which we detail below.

Analyzing chamber interactions—based on shared information sources—offers a more robust and informative view of ideological alignment than direct audience analysis, due to higher signal-to-noise ratios, greater stability, and low correlation with audience size. By focusing on second-order connections, chamber overlaps reveal deeper ideological coherence, capturing shared influences even among users without direct ties or explicit labels (see Appendix~\ref{app:comparison} for supporting analyses).

Our approach to measuring overlaps draws on principles similar to co-citation projections in bibliometrics \cite{newman2018networks, ahlgren2003requirements, rodriguez2015new, malliaros2013clustering}. In a co-citation network, an edge between users $i$ and $j$ represents the number of shared audience members they have. For example, Beguerisse et al. \cite{beguerisse2017and} calculated authority scores for key users in retweet networks based on the eigenspectra of these co-citation networks, identifying influential agents in the social networks they analyzed. Similarly, Becatti et al. \cite{becatti2019extracting} applied this projection to bipartite retweet networks between verified and non-verified X users, revealing communities of verified users through audience similarities. Unlike co-citation or path-based similarity measures, which rely on undirected networks and require the full adjacency matrix \cite{leicht2006vertex, malliaros2013clustering}, chamber overlaps offer a scalable alternative. They capture second-order similarities directly from interaction patterns and are better suited for large, directed, and heavy-tailed networks, such as ours.

Following our analysis from Section \ref{sec:leading-users}, we expect the climate change discussion to exhibit a clear separation between climate believers and climate skeptics \cite{williams2015network, cody2015climate}. We quantify such a separation by looking at the aggregate chamber overlap distribution over the year; i.e., we consider $p(q)$ for $q \in (q_{ij}^t)_{t, i<j}$. Ideological separation implies significantly higher overlap within groups than between them. In its absence, overlaps remain well above zero--something our method can also detect.

\begin{figure}[ht!]
    \centering
    \includegraphics[width=0.7\textwidth]{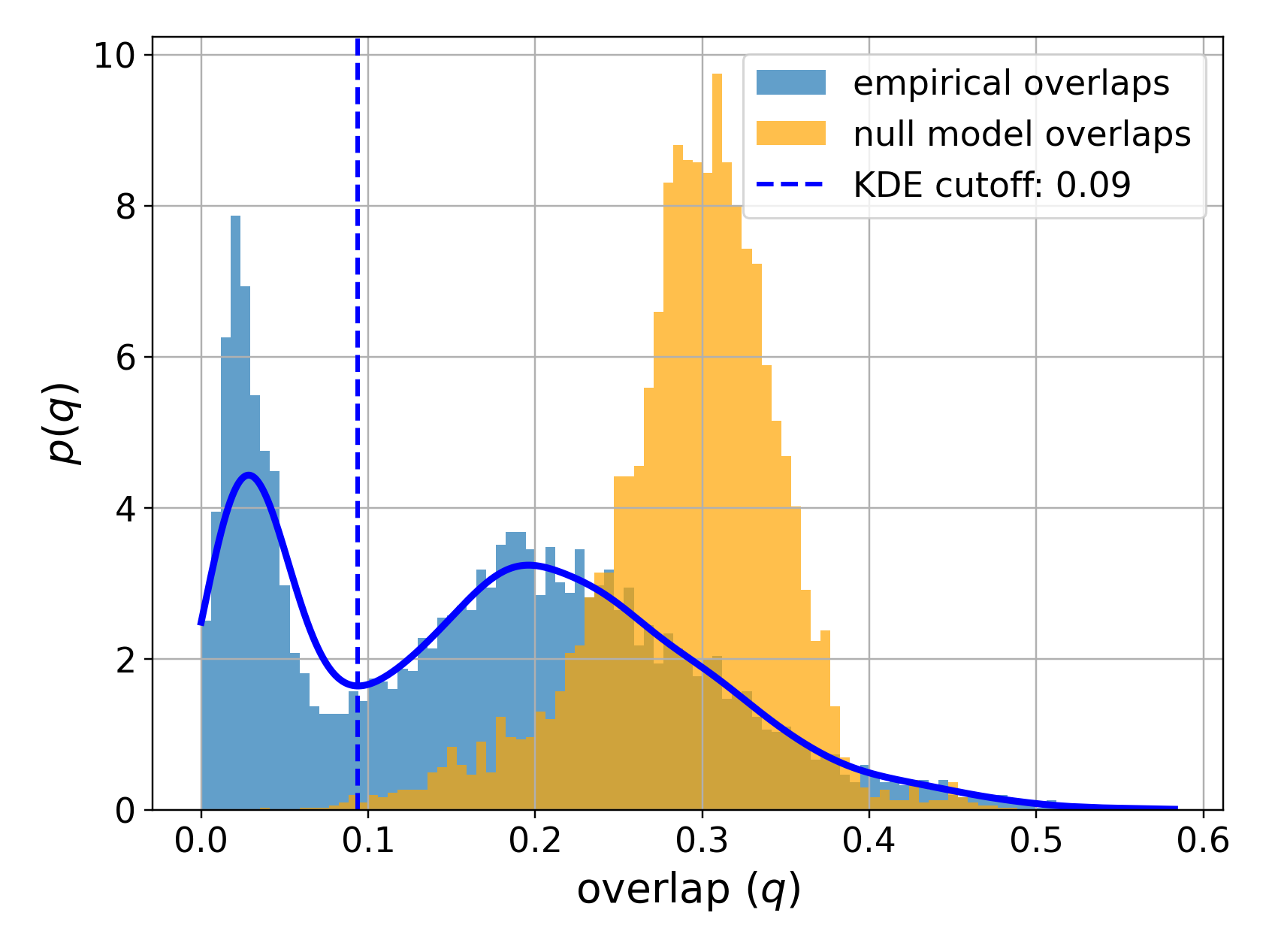}
    \caption{\textbf{Aggregate chamber overlap distributions for the empirical and configuration model networks.} We construct the aggregate empirical overlap distribution (blue) by concatenating the overlap pairs, $q_{ij}^t$, for every week $t$ on the dataset. We observe a \textit{bimodal} structure characterized by a low-overlap peak, $q_{off} = 0.04 \pm 0.02$, and a high-overlap peak, $q_{in} = 0.23 \pm 0.08$ (see the main text for details on how we separate the peaks).
    We compute the null-model chamber overlap distribution (orange) by averaging over 10 random network samples generated with the configuration model using the empirical degree sequences for each week. The resulting distribution is \textit{unimodal}, with a peak at $q_{\text{null}} = 0.29 \pm 0.05$.}
    \label{fig:overlap_dist}
\end{figure}

In Fig. \ref{fig:overlap_dist}, we show the overlap distribution, $p(q)$, aggregated over the whole period, where we observe a clear bimodal structure with a sharp peak at $q_{off} = 0.04 \pm 0.02$ and a more spread-out peak at $q_{in} = 0.23 \pm 0.08$. We approximate the position and spread of the peaks using a Gaussian kernel density estimator with an optimal bandwidth parameter, as determined by Scott's rule \cite{scott2015multivariate}, and separate the peaks based on the minimum between them. The observed bimodal structure clearly indicates the existence of distinct clusters within the leading users. The two peaks reflect the fact that given two random leading users, they may have a high overlap close to $q_{in}$ if they are in the same cluster, or a small overlap close to $q_{off}$ if they are in different clusters.

To compare, we constructed random graphs sampled from a configuration model with no difference between two leading users except their in-degree. The configuration model consists of a random graph ensemble where the edges are statistically independent but where the in-degree sequence is, on average, the same than the empirical network (see Appendix \ref{app:configuration} for details). For the configuration model, we find that the expected chamber overlap distribution is unimodal with a well-defined peak, which is consistent with the fact that there are no clusters between leading users by construction. We show the chamber overlap distribution for the empirical configuration model networks in Fig. \ref{fig:overlap_dist}.
The configuration model’s peak overlap is slightly higher than the empirical in-block value ($q_{null} = 0.29 \pm 0.05$), reflecting greater interconnectedness in the random, statistically unbiased network \cite{coolen2017generating} and suggesting a degree of randomness in the formation of the real network.
In Appendix \ref{app:configuration}, we derive explicit expressions for the expected chamber overlap, $\expected{q_{ij}^\chamber}$, and the expected audience overlap, $\expected{q_{ij}^\audience}$, between users $i$ and $j$ for a given in-degree sequence $\bk^{\textrm{in}}$ and out-degree sequence $\bk^{\textrm{out}}$:
\begin{align}
    \expected{q_{ij}^\chamber} &\approx \frac{\kin_i \kin_j }{N(\kin_i + \kin_j)} \frac{\expected{(\kout)^2}}{ {\expected{k}}^2} \frac{ \expected{(\kin)^2} } {\expected{k}} 
    \label{eq:null_chamber_overlap} , \\
    \expected{q_{ij}^\audience} & \approx \frac{\kin_i \kin_j }{N(\kin_i + \kin_j)} \frac{ \expected{(\kout)^2}}{ {\expected{k}}^2} .
    \label{eq:null_audience_overlap}
\end{align}
These expressions, asymptotic at $\mathcal{O}(1/N)$, give us a qualitative picture of how overlap depends on the moments of the degree distributions and the connectivity of the user pair. Importantly, chamber overlaps are amplified by the additional factor $\expected{(\kin)^2}/\expected{k}$, since $\expected{q_{ij}^\chamber} = \frac{\expected{(\kin)^2}}{\expected{k}} \cdot \expected{q_{ij}^\audience}$. In online social networks, in-degree (retweets received) is much broader than out-degree (retweets given), making expected chamber overlaps substantially greater than audience overlaps.

The overlap distribution reveals polarization in the climate conversation, but does not indicate the number of groups or assign users to specific groups. In what follows, we describe how to obtain the ideological position of the users based on their chamber overlap, $q_{ij}^t$, and compare it quantitatively with manual labeling. 

\subsection{Classifying leading users by chamber overlap}
 
In this section, we create a partition of the set of \emph{all} leading users, $\topusers^\Delta$, based on the chamber overlap distribution, $p(q)$, which exhibits a bimodal structure (See Fig. \ref{fig:overlap_dist}). A partition corresponds to a collection of disjoint subsets of $\topusers^\Delta$, $\calP = \{P_\alpha \}_{\alpha \in I }$ with $P_\alpha \subseteq \topusers^\Delta$ and $\cup_{\alpha \in I} \calP_\alpha = \topusers^\Delta$. In general, we could create a partition with an arbitrary number of clusters. Still, given that the X climate discussion exhibits a natural ideological division into climate believers and climate skeptics, we expect a separation of $\topusers^\Delta$ into two clusters.

To create a partition of the leading users, we first consider the (temporal) chamber overlap matrices between all pairs of leading users, $\bm{Q}^t = (q_{ij}^t)_{i,j \in \topusers^\Delta(t)}$. By treating $\bm{Q}^t$ as weighted, undirected, adjacency matrices, we can dispose of a plethora of \textit{community detection} algorithms to detect communities in networks. Thus, we classify the leading users \textit{in an unsupervised way} by first considering the \textit{aggregate} chamber overlap matrix, $\bm{Q}:= \expected{\bm{Q}^t}_t$, where $\expected{\cdot}_t$ denotes average over $t$, and then partitioning $\bm{Q}$ into \textit{two groups} using the spectral clustering algorithm described by Mohar et al. \cite{mohar1997some} (see Appendix \ref{app:unsupervised_communities} for details). We choose a spectral clustering algorithm because 1) it naturally partitions the network into two distinct groups, and 2) it ranks the nodes in the network (here, the leading users) according to how well-separated they are from the out-group. We assign labels based on observed ideology, but since clustering is fully unsupervised, we refer to the approach as unsupervised with only weak supervision at the cluster labeling stage. In general, we could consider other community detection algorithms \cite{blondel2008fast, rosvall2008maps, peixoto2019bayesian} if we need to partition the network into more than two groups. 

\begin{figure}[ht!]
	\begin{subfigure}[b]{0.59\textwidth}	
	\includegraphics[width=\textwidth]{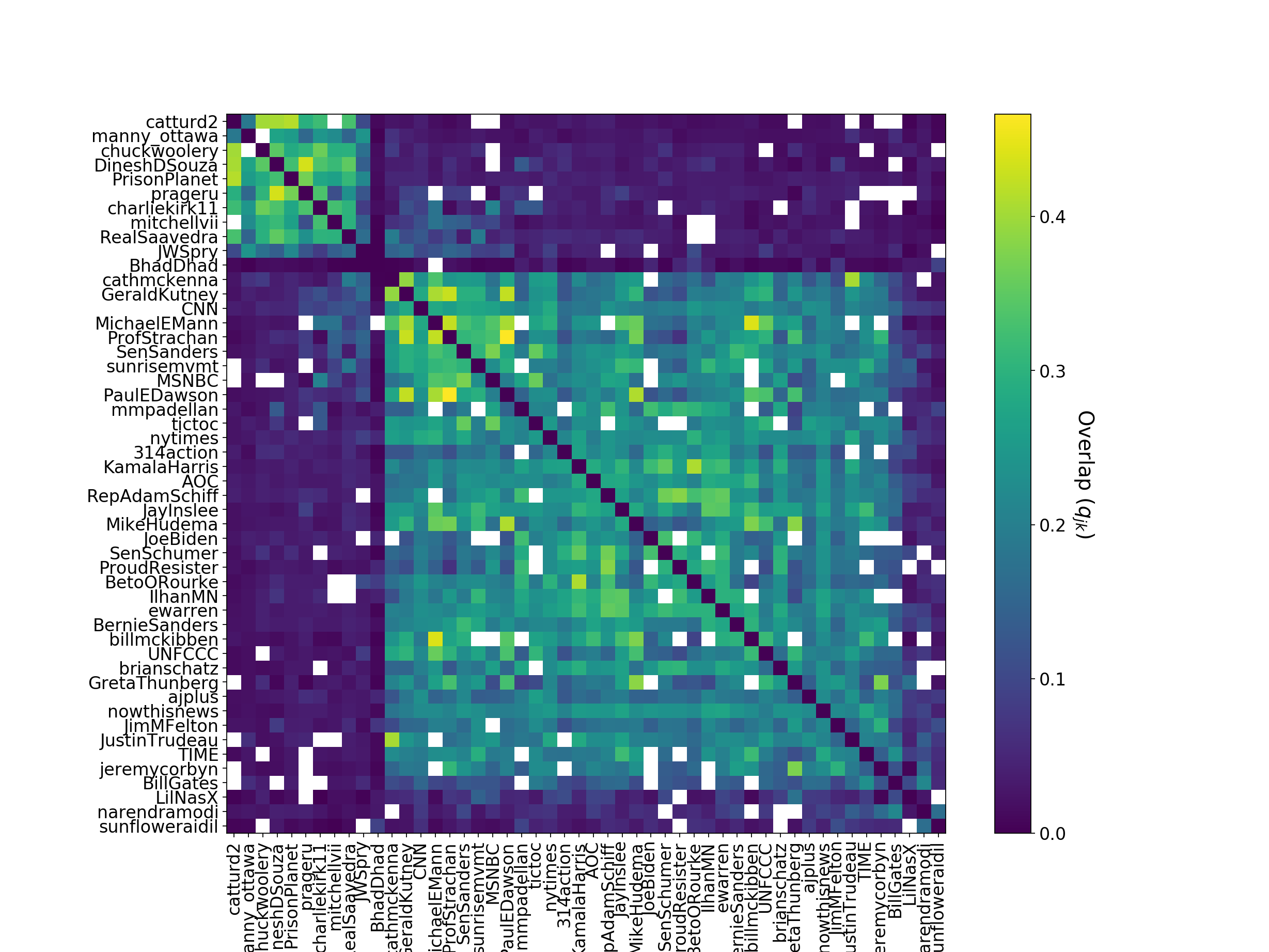}
	\caption{ Aggregate chamber overlap matrix }	
	\label{fig:aggregate_overlap_similarities_matrix}
	\end{subfigure}
	\centering
	\begin{subfigure}[b]{0.4\textwidth}
	\includegraphics[width=\textwidth]{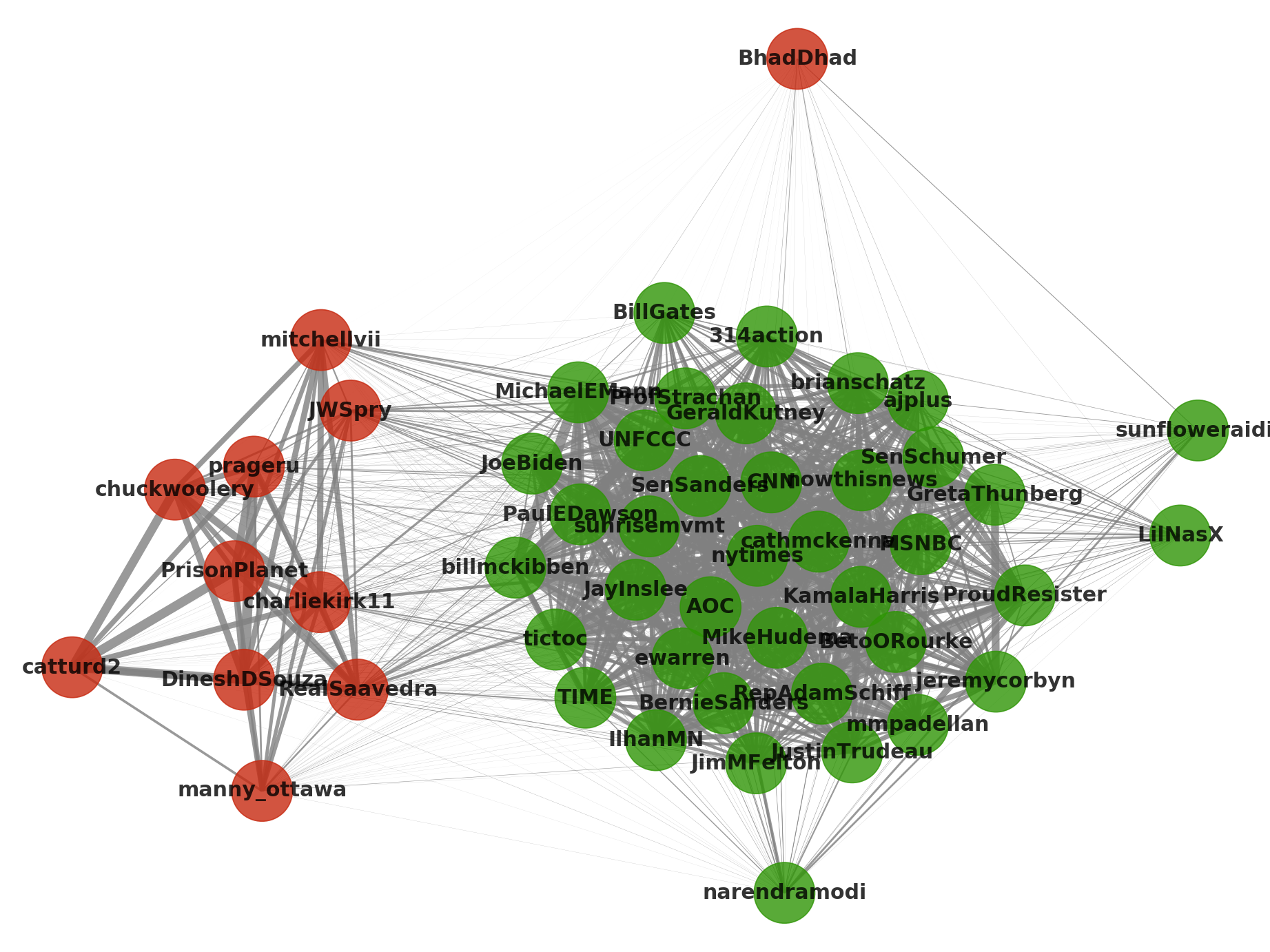}
	\caption{ Aggregate similarity network }	
	\label{fig:aggregate_overlap_similarities_network}
	\end{subfigure}
	\caption{\textbf{Aggregate chamber overlaps between the leading users}. \textit{a)} Aggregate chamber overlap matrix, $\bm{Q}$, of every leading user pair (see main text for details). White pixels represent leading users that were never present simultaneously during the same week. We order the users in $\bm{Q}$ according to the rank we obtain with the \textit{unsupervised} spectral clustering algorithm (see Appendix \ref{app:unsupervised_communities} for details). \textit{b)} Weighted similarity network constructed from $\bm{Q}$. We color the nodes according to the partition obtained by unsupervised spectral clustering. We identify two groups: \textit{climate believers} (green) and \textit{climate skeptics} (red), based on the users' profiles within each group.}
	\label{fig:aggregate_overlap_similarities}    
\end{figure}

In Fig. \ref{fig:aggregate_overlap_similarities_matrix}, we present the aggregate chamber overlap matrix, $\bm{Q}$, where we sort the users according to the rank given by the spectral clustering algorithm (see Fig. \ref{fig:eigenvector} in Appendix \ref{app:unsupervised_communities}). We observe a clear block structure that separates the climate skeptics (top left) from the climate believers (bottom right). Notice that the in-block overlaps are often two-digit percentages: a large fraction of the chamber is shared by users in the same community. Moreover, in Fig. \ref{fig:aggregate_overlap_similarities_network}, we show the resulting network constructed with $\bm{Q}$, where the node colors represent the partition found by the spectral clustering algorithm. We observe two well-separated groups that correspond to the climate believers and climate skeptics discussed before. When manually labeling users, we identified a third group that includes news channels and other media accounts. Spectral clustering only produces two classes. However, accounts manually labeled as neutral—namely \textit{CNN, NYTimes, ajplus}, and \textit{nowthisnews}—are statistically indistinguishable from the climate believers group, so spectral clustering assigns them to the climate believers cluster. Given this statistical similarity, other clustering algorithms would likely also assign these news media accounts to the climate believers. Ignoring neutral users, unsupervised clustering achieved perfect accuracy in comparison to manual labeling. However, the leading users \textit{BhadDhad} and \textit{LilNasX} exhibit a significantly low overlap with every leading user, so another clustering algorithm would probably make isolated clusters for them. We find two other users with a significantly lower overlap than the rest of the in-group user pairs, namely, \textit{narendramodi} and \textit{sunfloweraidi}. In a few cases, there are cross-chamber overlaps that are higher than the average cross-chamber overlap, such as in the pair \textit{cathmckenna}(believer)-\textit{RealSaavedra}(skeptic). However, such cross-chamber overlaps are much smaller than those within chambers.

\subsubsection{Polarization dynamics of the leading users}\label{sec:polarization}

The previous analysis suggests a large ideological polarization in climate-related conversations on X, a well-documented phenomenon in discussions about climate change \cite{williams2015network, jang2015polarized, chen2021polarization}. However, our results are valuable in that 1) we detect these groups in an unsupervised way, and 2) we quantify the relative ideological similarity between leading users using the chamber overlaps, where we found significantly high (low) overlaps for pairs of users in the same (other) group. However, we have not yet quantified polarization, and the analysis has been static. 

We define the \textit{polarization} between two groups $P_\alpha$ and $P_\beta$ as the difference between the probability that an edge exists within a group and the probability that an edge exists across groups. To do so, we consider the \textit{adaptive E-I index}, following Chen et al. \cite{chen2021polarization} and Bruns \cite{bruns2017echo}, as
\begin{equation}
    \Phi(t | P_\alpha, P_\beta) = \frac{n_{\alpha\alpha}^t + n_{\beta\beta}^t - ( n_{\alpha\beta}^t + n_{\beta\alpha}^t )}{n_{\alpha\alpha}^t + n_{\beta\beta}^t + ( n_{\alpha\beta}^t + n_{\beta\alpha}^t )}, 
    \label{eq:polarisation}
\end{equation}
where $n_{\alpha \beta}^t = \sum_{i \in P_\alpha, j \in P_\beta} q_{ij}^t $ is the total strength of the edges going from $P_\alpha$ to $P_\beta$. 

Our measure of polarization is bounded such that $\Phi \in [-1,1]$, where $\Phi = 1$ when all the connections happen within groups (total assortativity), $\Phi = -1$ when all connections happen between groups (total disassortativity), and $\Phi = 0$ when the connections between groups equal the connection within groups (no assortativity). We remark that Eq. (\ref{eq:polarisation}) is not the only way to quantify polarization. Most operational definitions of polarization involve comparing the out-group against the in-group interactions. See Table \ref{tab:polarization_desc} in Appendix \ref{app:echochamber_desc}, where we present different notions of polarization and their mathematical quantification as described in the literature.

\begin{figure}[ht!]
    \centering
	\includegraphics[width=0.7\textwidth]{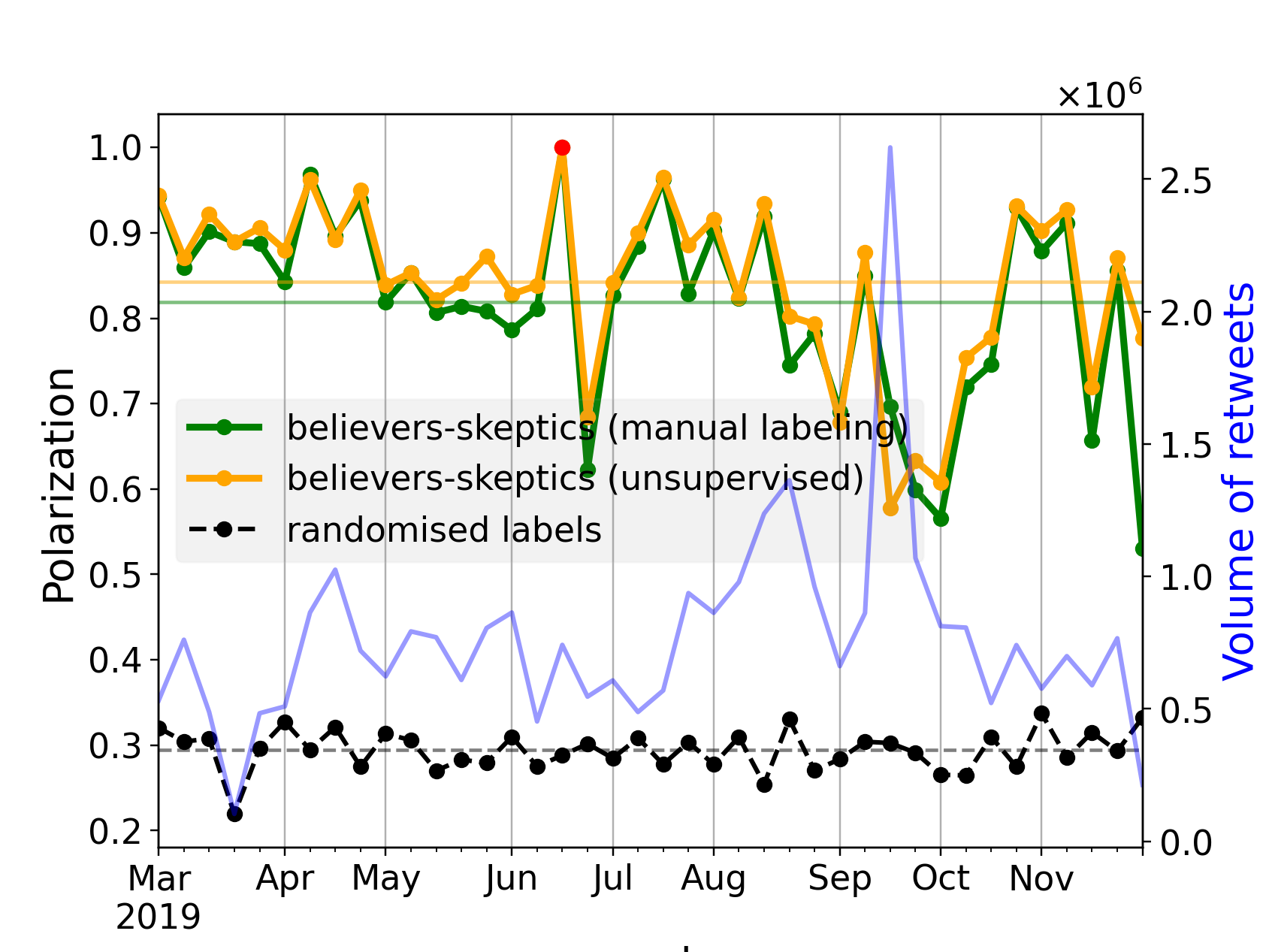}
	\caption{\textbf{Weekly polarization dynamics} of the leading user overlap matrices, $\bm{Q}^t$, (see Eq. (\ref{eq:jaccard})) for the manual labeling of climate believers and skeptics (green, $\langle \Phi_{\text{manual}} \rangle = 0.83$), the unsupervised partition using spectral clustering (orange, $\langle \Phi_{\text{unsup}} \rangle = 0.84$) and the average over an ensemble of randomly reshuffled labels over the same networks (black, $\langle \Phi_{\text{null}} \rangle = 0.23$). Red markers indicate weeks without climate skeptics. To provide a sense of scale, we display the total number of retweets per week in blue.}
	\label{fig:polarization_dynamics}
\end{figure}

In Fig. \ref{fig:polarization_dynamics}, we show the polarization dynamics of the leading users for the manual labeling and unsupervised clustering of climate \textit{believers} and \textit{skeptics} (see Fig. \ref{fig:impact_dynamics}). As a null model benchmark, we compute the polarization dynamics of the average over an ensemble of networks with randomly shuffled labels. 

We find a high polarization of $\Phi_{\text{manual}} = 0.83 \pm 0.11$ and $\Phi_{\text{unsup}} = 0.84 \pm 0.11$ for manual labeling and unsupervised clustering, respectively, which means that users with similar ideologies have similar chambers (climate-related). Moreover, we show that the polarization dynamics between manual labeling and unsupervised clustering are almost identical. Even when we tried to separate news media sources from climate believers, unsupervised clustering placed them in the same group. We find an average polarization of $\Phi_{\text{null}} = 0.23$ over $100$ realizations of randomly reshuffling the group labels. This low value indicates that neither the degree sequence nor the size of each ideological group can account for the high polarization in the empirical network. We find no significant temporal trends in the polarization dynamics but observe that polarization decreases during the 3$^{rd}$ and 4$^{th}$ week of September, which coincides with the \#FridaysForFuture largest strikes organized by Greta Thunberg \cite{laville2019across, taylor2019climate}. We discuss this and other signals that coincide with strikes later in Section \ref{sec:augmented_echo_chambers}.

\section{From chambers to echo chambers}\label{sec:echo_chambers}

Previously, we introduced the \textit{chamber} as the many-to-many information source of the \textit{audience} of a \textit{leading user}. In the case of the climate change retweet network, we found that the overlap distribution between all chamber pairs is bimodal (see Fig. \ref{fig:overlap_dist}), suggesting that such communication channels are divided, polarizing the climate change retweet network. Thus, following an unsupervised clustering approach, we classified the chambers according to their overlap similarities. We found two well-separated groups, which we identified as climate believers and climate skeptics.  

While many authors have studied polarization and echo chambers in social networks, there is still no consensus on what an echo chamber is. However, some characteristics of echo chambers are transversal to all its definitions: homophilic interactions drive their formation \cite{xia2021spread, gillani2018me, cinelli2021echo}, where actors in the system choose to connect while simultaneously excluding outsiders preferentially \cite{bruns2017echo}, and attitudes and beliefs stay inside groups of like-minded people \cite{garimella2018quantifying}. In particular, for interaction networks in social media, we assume that information flows through the edges of the network and that bimodal structures indicate the presence of echo chambers \cite{cinelli2021echo}.

Under this framework, the leading users associated with an ideological group (e.g. climate skeptics), together with their audiences and their chambers, \emph{approximate} an \textit{echo chamber}. Our rationale is that information flows in the opposite direction of retweets, as we schematize in Fig. \ref{fig:diagram_chamber}, and such information flows mostly between high-overlapping chambers. The overlap between the two chambers indicates the proportion of users they have in common. Therefore, a multimodal overlap distribution with high-overlap modes and a low-overlap mode (close to 0) suggests that users choose to preferentially connect (high-overlap modes) \textit{and} exclude outsiders (low-overlap mode). If a network exhibits such characteristics, the set of leading users should be easily partitioned as described in the previous section; thus, we introduce a definition of an echo chamber.

\begin{figure}[ht!]
    \centering
	\includegraphics[width=1\textwidth]{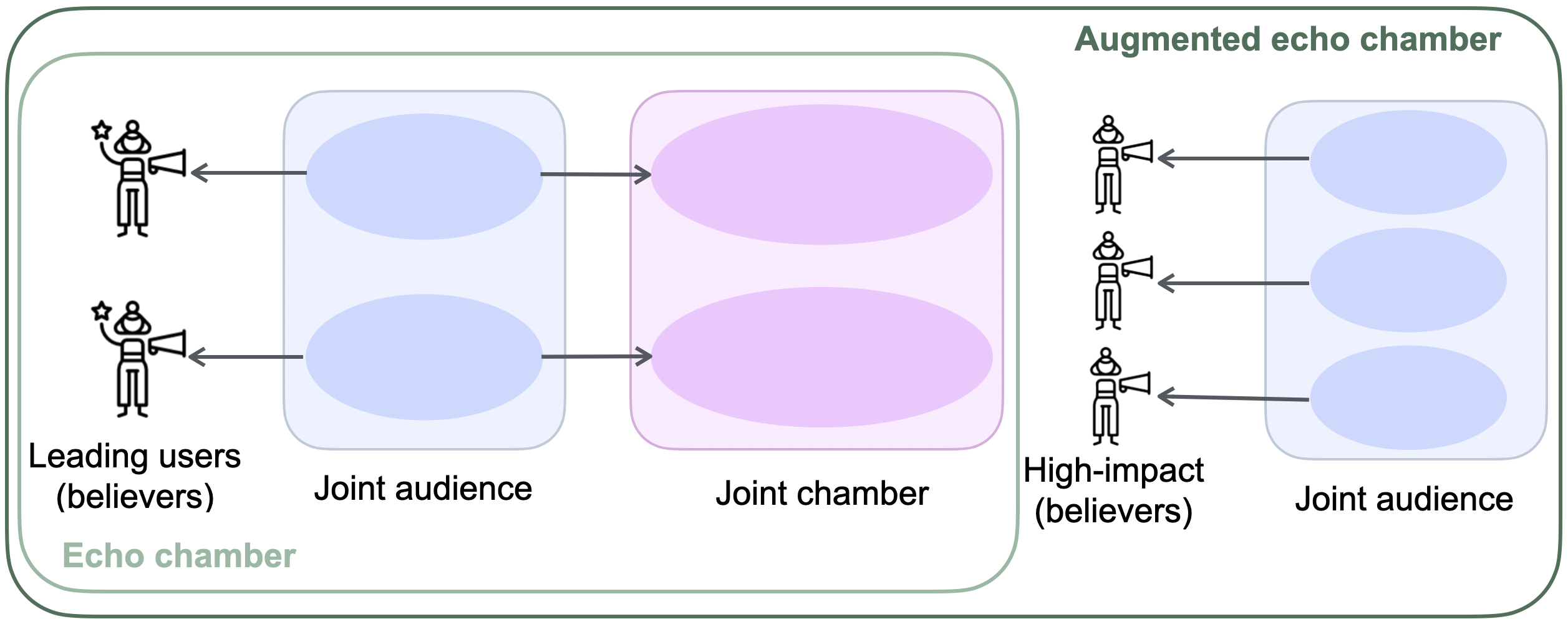}
	\caption{\textbf{Schematic of an echo chamber and an augmented echo chamber.} An echo chamber (Def.~\ref{def:echo-chamber}) includes leading users classified with the same ideology (e.g., climate believers), along with their audiences and chambers. An augmented echo chamber (Def.~\ref{def:augmented-echo-chamber}) extends this by incorporating high-impact users with the same ideology and their corresponding audiences (as defined in Eq. (\ref{eq:ideology_score})).}
        \label{fig:echo_chamber_diagram}        
\end{figure}

\begin{definition}[Echo chamber\label{def:echo-chamber}]
Given a partition of the leading users, $\topusers^\Delta = \cup_{x \in I} \calP_x$, obtained by clustering the chamber overlap matrix, $\bm{Q} = (q_{ij})_{ij \in \topusers^\Delta}$, where the overlap distribution $p(q)$ exhibits a clear peak near $q = 0$ and other distinct peak(s) far from $q=0$, we define the \textit{echo chamber} of a group $\calP_x$ a the union of the leading users, chambers, and audiences associated with $\calP_x$. Mathematically, 
\begin{equation}
    \mathcal{E}_{\calP_x} = \bigcup\limits_{i \in \calP_x} \left( \{i\} \cup \calA_i \cup \calC_i \right)
    \label{eq:echo_chamber}
\end{equation}
where $\calA_i$ is the audience of $i$ and $\calC_i$ her chamber. 
\end{definition}
An echo chamber is an ill-defined concept; therefore, we note that Definition \ref{def:echo-chamber} provides an approximation that works well for retweet networks with multimodal overlap distributions, $p(q)$. Figure~\ref{fig:echo_chamber_diagram} illustrates the echo chamber defined in Definition~\ref{def:echo-chamber}, along with a preview of augmented echo chambers, which we formally introduce in the next section. In this work, we only consider partitioning into two groups, but the approach can be generalized using other clustering algorithms. 

Structurally, the climate change retweet network is well-separated, so we claim that we can observe ideological echo chambers: climate believers and climate skeptics, which we abbreviate as $B$ and $S$, respectively. We construct two dynamic echo chambers per week based on the leading users, $\topusers^\Delta(t)$, and our partition of climate believers and skeptics. The echo chamber for believers, $\mathcal{E}_B$, contains $14.8 \pm 7.8 \%$ of the total users per week, while the echo chamber for skeptics, $\mathcal{E}_S$, contains $2.6 \pm 2.1 \%$. Together, they cover $17.4 \pm 9.9 \%$ of the total retweeting population per week. These echo chambers may have some users that overlap by construction, that is, users classified as both believers and skeptics. However, we find that only $0.3 \pm 0.1 \%$ users are in both echo chambers, indicating that the cross-communication between them is orders of magnitude lower\footnote{It might be the case that if we inspected the reply network or the followers' network instead of the retweet network, the intensity of cross-communication between echo chambers could be higher.}.

\subsection{Augmented echo chambers}\label{sec:augmented_echo_chambers}

In the previous section, we identified two echo chambers: the climate believers chamber ($\mathcal{E}_B$) and the climate skeptics chamber ($\mathcal{E}_S$). Together, these represent only a small portion of the retweeting population ($17.4 \pm 9.9 \%$), since they consider only the leading users, $\topusers^\Delta(t)$, and exclude the broader set of high-impact users, $\topusers(t)$. To further explore ideological alignment, we analyze the audiences of other high-impact users and assess whether they lean toward either ideological group based on $\mathcal{E}_B$ and $\mathcal{E}_S$.

To do this, we introduce an \textit{ideology score}, similar to that in \cite{del2017mapping}, which measures whether the audience of a high-impact user is predominantly composed of believers, skeptics, or neither. We compute the ideology score $s_i$ of high-impact user $i$ as
 \begin{equation}
    s_i = \frac{ n_B^i - n_S^i }{ n_B^i + n_S^i },
    \label{eq:ideology_score}
\end{equation}
where $n_X^i = | \calA_i \cap \mathcal{E}_X |$ is the number of users in $i$'s audience who are also in the echo chamber $\mathcal{E}_X$. The score $s_i$ ranges from $-1$ when every user in $\calA_i$ is a climate skeptic to $1$ when all are climate believers.

When the ideology score has a large magnitude, we can associate them and their audience with an ideological group, based on the sign of $s_i$: we identify users with scores near $-1$ as skeptics and those near $1$ as believers. Using this approach, we can \textit{augment} the echo chambers by including the audiences of high-impact users with strong ideological scores, as detailed in the following definition.
\begin{figure}[ht!]
    \centering
    \includegraphics[width=0.7\linewidth]{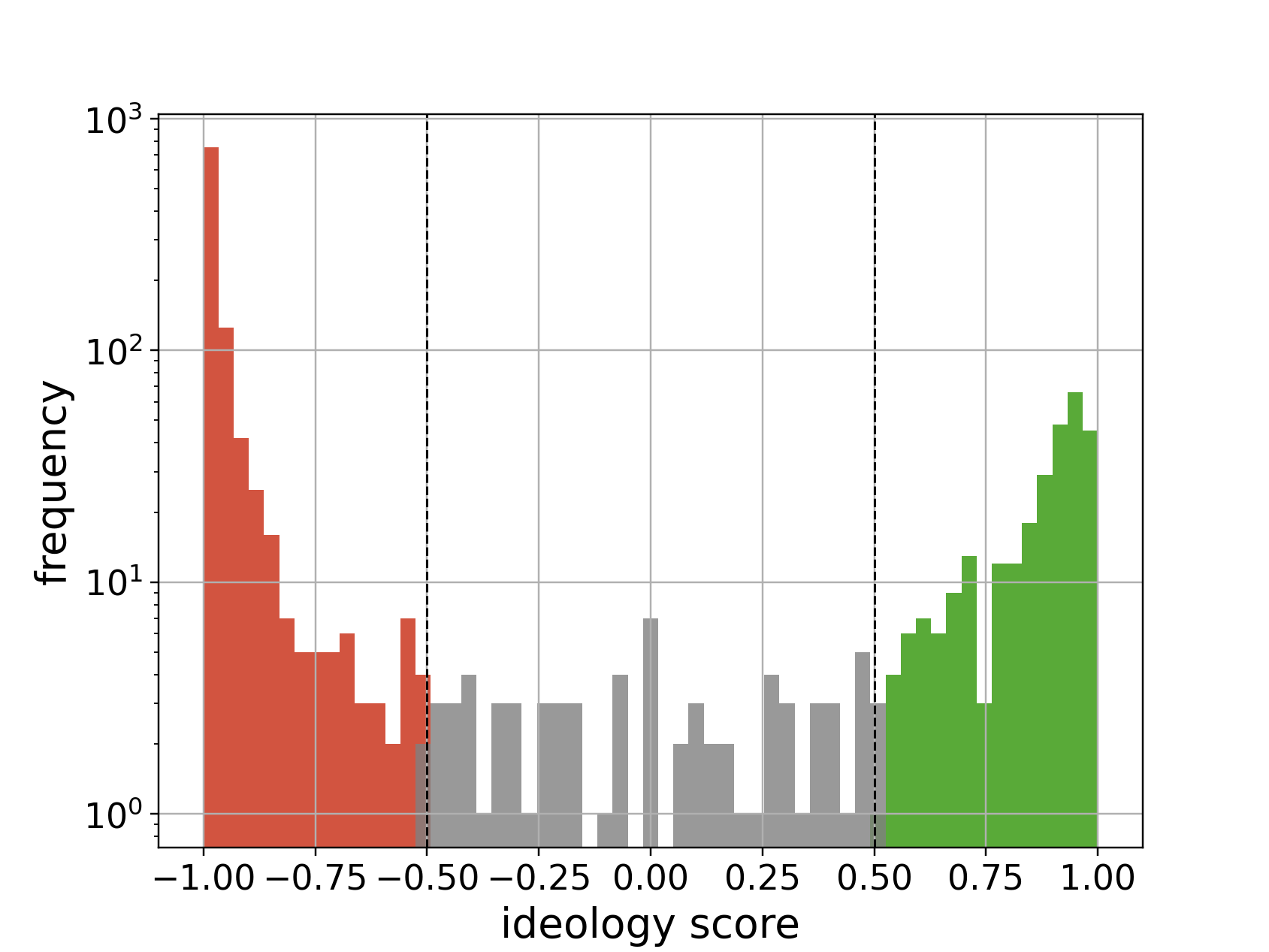}
    \caption{\textbf{Ideology scores distribution} for the high-impact users, $\topusers$, where we find a clear bimodal structure (see Eq. (\ref{eq:ideology_score})). We choose a threshold $\eta = 0.5$ for which users with $s_i > \eta$ (green) are assigned as climate believers and those with $-s_i > \eta$ (red) are assigned as climate skeptics. The augmented echo chambers do not include users with intermediate scores ($s_i \in [-\eta, \eta]$).}
    \label{fig:ideology_scores}    
\end{figure}

\begin{definition}[Augmented echo chamber\label{def:augmented-echo-chamber}]
Given the echo chambers $\mathcal{E}_B$ and $\mathcal{E}_S$ and the collection of high-impact users for which $| s_i | \geq \eta$, with $\eta$ a user-defined threshold, we define their corresponding \textit{augmented echo chambers}, $\mathcal{E}_B^a$ and $\mathcal{E}_S^a$, as follows 
\begin{align}
    \mathcal{E}_B^a &= \bigcup\limits_{i: s_i \geq \eta} \left( \{i\} \cup \calA_i \right) \cup \mathcal{E}_B , \\
    \mathcal{E}_S^a &= \bigcup\limits_{i: -s_i \geq \eta} \left( \{i\} \cup \calA_i \right) \cup \mathcal{E}_S ,
\end{align}
where $\calA_i$ is the audience of $i$.
\end{definition}
This is one of many ways to define an echo chamber (see Figure \ref{fig:echo_chamber_diagram} for a schematic illustration of augmented echo chambers). The literature offers various perspectives on polarization and echo chambers—see Tables \ref{tab:polarization_desc} and \ref{tab:echochamber_desc} in Appendix \ref{app:echochamber_desc} for a summary of these different definitions.

In Fig. \ref{fig:ideology_scores}, we show the ideology score distribution for the remaining high-impact users, i.e., those who are not leading users. The distribution is clearly bimodal, with the mode corresponding to climate believers (right) being significantly higher than that of climate skeptics (left). We choose $\eta = 0.5$ as our threshold to decide whether a high-impact user is included in the augmented echo chamber. Using this method, of the $1360$ remaining high-impact users across all weeks, we classified $280$ as climate skeptics, $1078$ as climate believers, and $73$ users remain unclassified. We validated this classification by manually labeling a random sample of 100 high-impact users, achieving remarkable performance (Acc. $0.99$, F1 $0.97$; see Appendix \ref{app:manual_labeling} for details). These results support the reliability of our definition of augmented echo chambers.

In addition to augmenting the number of users in the echo chambers, this method enables us to discover the ideological position of high-impact users in an unsupervised way. For example, we identify users like \textit{@RollingStone} ($s_i = 0.989$) or \textit{@Greenpeace} ($s_i = 0.995$) as climate believers while users like \textit{@DonaldJTrumpJr} ($s_i = -0.93$) or \textit{@GovMikeHuckabee} ($s_i = -0.948$) as climate skeptics solely based on their scores. Although we do not thoroughly analyze unclassified users, we noticed that low-scoring users tend to stay neutral on climate change, for example, \textit{@SkyNews} ($s_i = 0.17$), or are not a strong part of their content agenda, e.g., \textit{@Imamofpeace} ($s_i = 0.03$). Beyond these examples, we manually labeled $100$ randomly selected non-high-impact users from the augmented echo chambers and observed very strong classification performance (Acc. $0.95$, F1 $0.96$; see Appendix \ref{app:manual_labeling} for details). This further validates our approach, showing that it can reliably assign ideological labels to users beyond those with high impact, enabling scalable and accurate classification across the broader population.

\begin{figure}[ht!]
    \centering
    \includegraphics[width=0.8\linewidth]{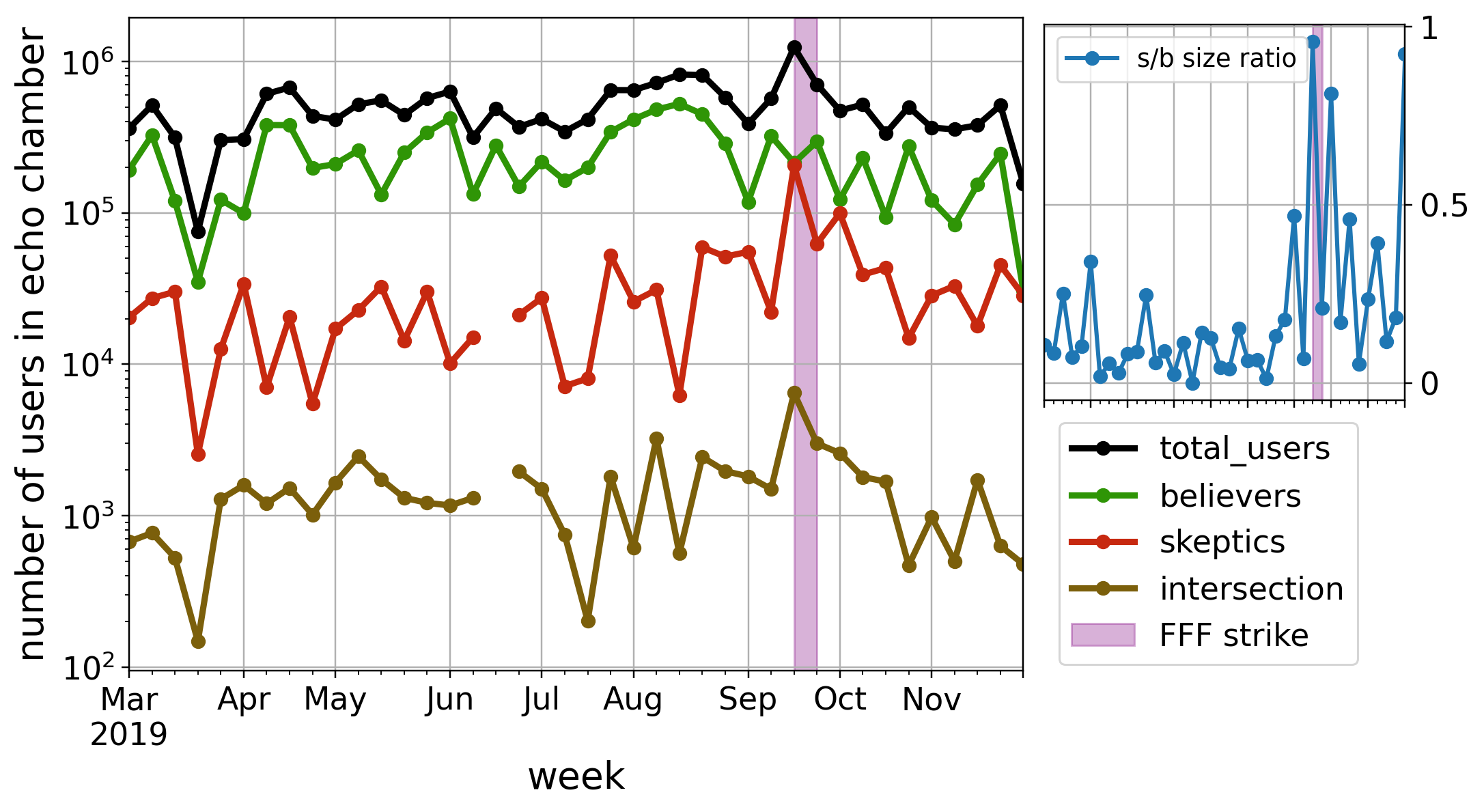}
    \caption{\textbf{Size of the augmented echo chambers} in logarithmic scale for every week in the dataset. In green and red, we show the number of believers and skeptics, while in brown, we show the number of users in the intersection. We omit a data point in June 2019 because the skeptics' echo chamber, $\mathcal{E}_S$, was empty during that week. In black, we show the total number of users in the retweet network for that week. The inset plot shows the size ratio of skeptics against believers, where we observe a significant increase during late September 2019, which we suspect was triggered by the Fridays for Future biggest strike \cite{laville2019across, taylor2019climate} (vertical pink interval).}
    \label{fig:echo_chamber_sizes}    
\end{figure}

With the augmented echo chambers, we classify $53.1 \pm 9.8 \%$ of the weekly total retweeting population. This result is consistent with our initial observation that the top $N=50$ high-impact users are retweeted by around $50\%$ of the total weekly users. Moreover, this augmentation increases the sizes of the original echo chambers by a factor of $3.5$. More specifically, the augmented echo chamber of believers is larger than its original counterpart by $3.7 \pm 1.6$, the skeptics by $2.6 \pm 1.2$, and the users at the intersection only by $1.4 \pm 0.4$. 

As validation, we benchmark our augmented echo chamber classification against Barbera's latent ideology inference model \cite{barbera2015birds},  a widely used method in political and climate communication research \cite{falkenberg2022growing}. Our classification achieves almost identical performance while covering nearly four times more users, thanks to its scalable use of chamber structures (see Appendix~\ref{app:latent_ideologies} for full results and implementation details).

In Fig. \ref{fig:echo_chamber_sizes}, we present the sizes of the augmented echo chambers, the number of users in their intersection, and the total number of retweeting users each week. These sizes are stable for most of the year, with believers oscillating around $2.4 \times 10^5$ users and those at the intersection oscillating around $1500$ users. In contrast, the number of climate skeptics rises significantly from before the Fridays for Future September strikes \cite{laville2019across, taylor2019climate}, oscillating from around $2.4 \times 10^4$ users to oscillating around $5.6 \times 10^4$ users after the strikes. In the inset of the figure, we show the ratio of skeptics to believers, which consistently increases after the strikes. This is counterintuitive, given that the strikes were pro-climate events.

We believe the correlation between climate strikes and the increase in the skeptics' impact is not random. Besides such an increase, we already observed in Figure \ref{fig:polarization_dynamics} a significant drop in polarization at the dates of the strikes that were not caused by changes in chamber sizes. There are several explanations for this correlation with size and polarization, ranging from an endogenous social reaction among climate skeptics to a coordinated response using bots \cite{cinelli2022coordinated}, where the bots massively retweeted the skeptic \textit{and} believers leading users during the dates of the strikes. While we cannot determine its causes with this analysis, we find it remarkable that we can observe such a signal from several angles, purely through our unsupervised construction of echo chambers. 

\subsection{Renewal of users within ideological groups}\label{sec:autooverlap_flow}

Most of the statistical features in our analysis remain stable over time, ranging from the bimodal structure of the overlap distribution to the number of users per week (except for the peak in September, which coincides with the Fridays for Future strikes) to the Gini index of users' impact. However, X (formerly Twitter) is very dynamic, with users entering and leaving the conversation as different topics emerge and decay over time \cite{comito2019bursty}, as is the case of the climate change conversation \cite{dahal2019topic}.

\begin{figure}[ht!]
    \centering
    \includegraphics[width=0.7\linewidth]{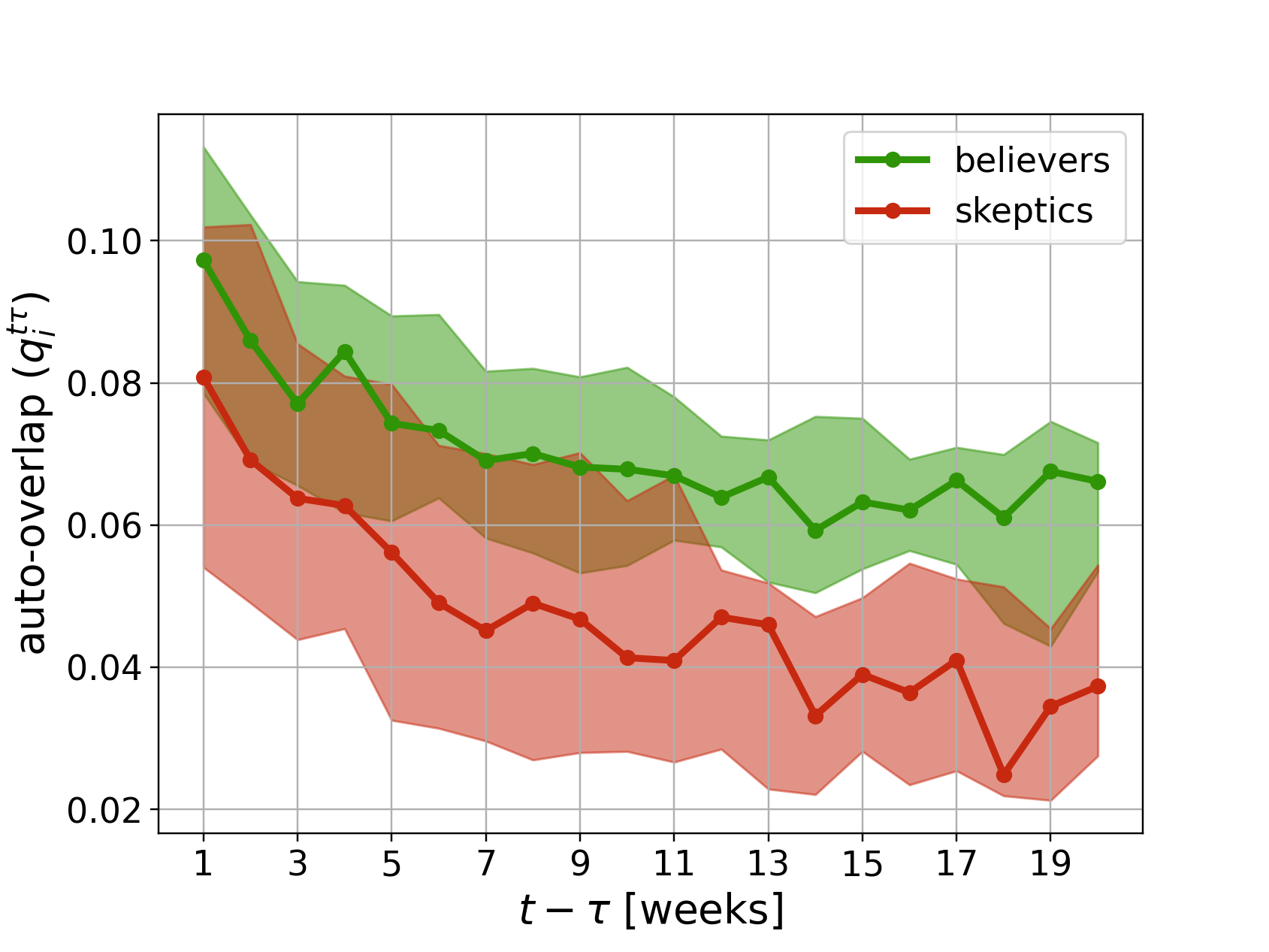}
    \caption{\textbf{Auto-overlap similarity decay for each augmented echo chamber} as a function of the time difference (in weeks) between them. The solid lines with markers indicate the median overlap similarity for each week, while the spread corresponds to the $25-75$ quantile spread.}
    \label{fig:echo_chambers_population_decay}    
\end{figure}

To understand whether the same users drive the climate conversation year-round or if there is a continuous renewal of participants who behave consistently on a broader scale, we examine the flow of users within augmented echo chambers over time. If there is little change in the user base, this would support the idea that the same people maintain the conversation. Conversely, a high turnover would suggest a continuous influx and exit of users. We measure this turnover by calculating the \textit{auto-overlap} of an augmented echo chamber, $\mathcal{E}_X^a(t)$, for each ideological group $X$ between two different weeks $t$ and $\tau$ as follows:
\begin{equation}
    q_{X}^{t\tau} = \frac{ \abs{\mathcal{E}_X^a(t) \cap \mathcal{E}_X^a(\tau)} } {\abs{\mathcal{E}_X^a(t) \cup \mathcal{E}_X^a(\tau)} } \ .
    \label{eq:auto_overlap}
\end{equation}
This measure is similar to the chamber overlap in Eq. (\ref{eq:jaccard}) but instead compares the augmented echo chambers for the same group at different times. A \textit{low} value of $q_X^{t\tau}$ indicates a \textit{high} turnover, and vice-versa. Importantly, this measure captures users who may leave the conversation temporarily and return later, allowing us to detect any cyclical patterns in user engagement.

In Fig. \ref{fig:echo_chambers_population_decay}, we show how the auto-overlap decreases over increasing weeks for each group. We find that climate believers retain more users over time than climate skeptics. Interestingly, only around $9.7 \pm 1.7 \%$ of believers and $8.1 \pm 2.4\%$ of skeptics remain in the conversation from one week to the next, reflecting significant user turnover within both groups. This high turnover aligns with our second hypothesis, suggesting that although users frequently enter and exit the conversation each week, the broader structure of the interaction networks remains stable.

\section{Conclusion}\label{sec:conclusions}

In this paper, we identified echo chambers in X’s (formerly Twitter) climate change conversation using unsupervised second-neighbor-based methods. Given that retweets on X typically concentrate around a few key users \cite{glenski2018propagation} and can serve as a proxy for endorsement \cite{barbera2015birds, gaumont2018reconstruction, chen2021polarization}, we analyzed the temporal structure of climate-related retweet networks to assess ideological similarities between leading users. We defined a leading user as one who consistently ranks among the most retweeted over multiple weeks.

We introduced the concept of a leading user's \textit{chamber}, defined as the set of users retweeted by that user's audience. Chambers represent many-to-many information sources linked to a leading user's audience \cite{liang2019did}, allowing us to measure ideological similarities between leading users based on their chamber overlaps. This overlap distribution was clearly bimodal, with peaks near zero (indicating minimal connectivity between groups) and further from zero (indicating strong connectivity within groups). For less polarized discussions, we do not expect overlaps to be near zero, as shown when reshuffling edges using the configuration model. Using an unsupervised spectral clustering algorithm \cite{mohar1997some}, we classified leading users into two groups: \textit{climate believers} and \textit{climate skeptics}, in line with previous findings \cite{williams2015network, tyagi2020polarizing, chen2021polarization}. We have found validation of our method in climate-related data beyond 2019 \cite{bassolas2024cross}, where authors also find a clear bimodal distribution representing climate believers and skeptics.

Based on the bimodal structure of the chamber overlap distribution, we defined an \textit{echo chamber} as the union of the same group's leading users with their associated audiences and chambers. This bimodality reflects key features of echo chambers: their formation is driven by homophilic interactions \cite{xia2021spread, gillani2018me, cinelli2021echo}, actors inside them preferentially connect while excluding outsiders \cite{bruns2017echo}, and attitudes and beliefs circulate within like-minded groups \cite{garimella2018quantifying}. We identified, on average per week, $15\%$ of the total retweeting population as climate believers, $3\%$ as climate skeptics, and only $0.3\%$ as belonging to both groups, indicating negligible cross-communication between echo chambers.

Furthermore, we designed an \textit{ideology score} that identifies the ideological position of any high-impact user, even if they have not been observed previously. This score depends on the proportion of her audience that belongs to either of the echo chambers. We found that the ideology score distribution over all the high-impact users is bimodal, with modes at opposite extremes of the ideological spectrum. This finding reinforces our claim that users within echo chambers are mostly like-minded. We validated the ideological positions of some high-impact users identified by their ideology score. For instance, we uncovered the global campaigning network Green Peace (\textit{@Greenpeace}) as a climate believer, the Republican governor Mike Huckabee (\textit{@GovMikeHuckabee}) as a climate skeptic, and the British news channel Sky News (\textit{@SkyNews}) as a neutral user.

Using the identified ideologies, we augmented our original echo chambers by incorporating the audiences of high-impact users who share the same stance , and confirmed the accuracy of this approach through manual labeling of a random user sample. These augmented echo chambers cover over half of the weekly retweeting population ($46.2\%$ climate believers, $6.5\%$ climate skeptics, and only $0.3\%$ overlapping users). Compared to the state of the art \cite{barbera2015birds, falkenberg2022growing}, our augmented echo chamber approach classifies nearly four times more users with similar accuracy and scalability.

Throughout most of the year, the sizes of each echo chamber remain relatively stable. However, we observe a strong positive correlation between the main \#FridaysForFuture strikes and growth in the skeptics’ echo chamber, although not among believers. We find it remarkable that we could identify the peak activity of certain parts of the population by using completely unsupervised methods. Additionally, we measured user flux within echo chambers over time and found that over $80\%$ of users leave their chambers from week to week, indicating that the stability of these echo chambers is an emergent property of the system rather than the result of a big group of users behaving consistently throughout the year.

We see several directions in which our chamber-based method can be improved. First, our analysis considered only two ideological groups, though X discussions might likely involve more groups. Future work could apply broader community detection algorithms, from Bayesian inference that lets you fix the number of groups \cite{peixoto2019bayesian} to those that optimize community structures with an unsupervised number of clusters \cite{blondel2008fast, rosvall2008maps}. The ideology score could be extended for multiple ideologies to a multi-class setting, such as measuring user proportions in each chamber within a higher-dimensional simplex.

Second, we did not differentiate user types, overlooking the possible impacts of verified accounts, non-human actors, and bots. Prior studies show that bots can amplify exposure to emotional content \cite{bessi2016social} and coordinate in controversial discussions \cite{cinelli2022coordinated, bruno2022brexit}. Integrating bot-detection methods could help create a user taxonomy to analyze these effects.

Finally, this approach could be applied to other platforms. Messaging apps like Telegram, where users interact in channels and groups \cite{bovet2022organization}, could benefit from chamber-based analysis by focusing on channel administrators. Due to its unsupervised nature, our approach could monitor large-scale polarization without directly examining content, preserving user privacy.

For our classification, we intentionally focused on \textit{who} X users engaged with rather than \textit{what} they consumed. While this approach lacks the detail of supervised methods, analyzing tweet content could further reveal users' ideological positions, reinforce evidence of echo chambers \cite{cinelli2021echo}, and uncover the main discussion topics within and between groups \cite{bovet2018validation}. However, our findings demonstrate that we can identify echo chamber structures even without examining tweet content. Although user ideology is complex and hard to determine at the individual level, we observe distinct, structurally cohesive ideologies at the group level. We also believe that using our method as an initial, unbiased analysis can be enriching before incorporating external information. Our unsupervised approach can highlight useful starting points for more complex, content-based analyses.

Overall, our work highlights the importance and usefulness of leveraging the structural information present in social media networks, especially when examining controversial conversations. Our methodology is computationally cheap, readily usable for other X datasets, and does not suffer from the selection bias of supervised approaches. Furthermore, if the conversation is polarized enough, we can identify echo chambers by looking at the handful of users who produce the most influential tweets, which is easier than deploying clustering algorithms on the entire network. From a social point of view, this condition shows that the climate-related X conversation-and possibly most conversations on X \cite{glenski2018propagation}-has very low complexity, meaning that we can identify large-scale structures within the conversation just from the activity of the few leading users.

\section*{Acknowledgments}
BK acknowledges funding from the Conacyt-SENER: Sustentabilidad Energ\'etica scholarship. FAL is supported by a grant from the Simons foundation (grant No. 454941, S. Franz).
The authors would like to thank Luca Mungo, Santiago Martínez Balvanera, Alexandre Bovet, and Doyne Farmer for proofreading the manuscript and providing very valuable feedback. BK is greateful with the INET Complexity Economics group for stimulating discussions. 

\section*{Code Availability}

All code used in this paper is written in \textit{Python} and is openly available in \url{https://github.com/blas-ko/Twitter_chambers}. The repository includes scripts to fully reproduce all the analyses and figures presented in the paper.

\printbibliography

\appendix
\section{Manual Labeling Protocol}\label{app:manual_labeling}
To validate our classification algorithm based on chamber similarities, we performed a manual labeling of three distinct user sets, assigning them an ideological stance as either \textit{climate believers} or \textit{climate skeptics}:

\begin{enumerate}
    \item The 50 leading users ($\topusers^\Delta$), 
    \item 100 randomly selected high-impact users ($\topusers$), and 
    \item 100 randomly selected users from the augmented echo chambers, balanced across ideological classes.
\end{enumerate}

The \textit{leading users} correspond to those analyzed in the main text and listed in Table~\ref{tab:leading users_description}. These are highly influential users and serve as reference points for comparison with our spectral clustering method.

The \textit{high-impact users} are influential accounts that, although not consistently active enough to be among the leading users, play a crucial role in expanding the reach of our classification. These users form the basis of the \textit{augmented echo chambers} described in Section~\ref{sec:augmented_echo_chambers}, enabling the ideological classification of approximately 53\% of the population, an increase from the 17\% directly classified via Eq.~\ref{eq:echo_chamber}.

The third set includes \textit{random users from the augmented echo chambers}—accounts that are not explicitly assigned an ideology score but are labeled based on their association with a high-impact user. This set allows us to test the validity of our inferred labels on indirectly classified users.

\vspace{0.5em}
\noindent\textbf{Labeling Procedure.} Authors BK and FA established the following protocol to ensure an unbiased and consistent annotation process:

\begin{itemize}
    \item Users were selected at random from each of the three predefined sets.
    \item Annotators were blind to the algorithmic classification before manual evaluation.
    \item Each annotator independently reviewed the user profile and timeline, examining approximately 100 tweets. If a user posted or retweeted content clearly aligned with either pro-climate or anti-climate positions, they were labeled accordingly.
    \item When explicit climate-related content was not present, an ideological stance was inferred from correlated views. For example, climate believers often express support for human rights, migration, reproductive rights, and progressive politics; skeptics tend to display pro-gun, religious, and conservative content \cite{kulin2021nationalist, hornsey2018relationships}. If such tendencies were consistently observed, the user was assigned a corresponding label.
    \item Only users for whom both annotators agreed were included in the evaluation. Accounts that were suspended, protected, or no longer active were excluded.
    \item A third label, \textit{other}, was used only for the leading users, to categorize public figures (e.g., artists or news outlets) with no discernible climate stance.
\end{itemize}

\noindent\textbf{Evaluation.}  
We treat the manual annotations as ground truth and evaluate the performance of our algorithm using standard metrics by assigning the positive class to \textit{climate believers} and the negative class to the \textit{climate skeptics}: true positives ($T_B$), false positives ($F_B$), true negatives ($T_S$), false negatives ($F_N$), accuracy, and F1 score. The results for each set of users are reported in Table~\ref{tab:manual_labeling_stats}.

\begin{table}[ht]
\centering
\caption{Manual labeling summary and algorithm performance}
\label{tab:manual_labeling_stats}
\begin{tabularx}{\textwidth}{lrrrrrrrrcc}
\toprule
\textbf{User set} & $n$ & $n_{\text{valid}}$ & $n_{B}$ & $n_{S}$ & $T_B$ & $F_B$ & $T_S$ & $F_S$ & Acc. & F1 \\
\midrule
Leading$^*$ & 50  & 41 & 31 & 10 & 31 & 0  & 10 & 0  & 1.0  & 1.0 \\
High-impact & 100 & 76 & 61 & 15 & 60 & 1  & 15 & 0  & 0.99 & 0.97 \\
Aug. Echo   & 100 & 41 & 20 & 21 & 18 & 2  & 21 & 0  & 0.95 & 0.96 \\
\bottomrule
\end{tabularx}
\end{table}

Across the three sets of users, the algorithm shows strong performance in all metrics. This indicates a high true positive and true negative rate, along with low false positive and false negative rates. In particular, consistent performance among users of the augmented echo chamber further supports the reliability of our expanded classification approach to capture the ideological stance of users.

\section{Comparison to latent ideologies inference method}\label{app:latent_ideologies}

We benchmark our augmented echo chamber (AEC) classification against Barbera’s latent ideology inference (LII) method \cite{barbera2015birds}, which infers user ideology from their direct interactions with high-impact users. We follow the implementation from Falkenberg et al. \cite{falkenberg2022growing}. Let $\mathbf{W}$ be the weighted adjacency matrix of interactions from audiences to high-impact users, pruned to users with at least two interactions, and let $m = \sum_{ij} W_{ij}$ be the total number of interactions. High-impact users, referred to as elites in the original method, are defined as the top $N$ users receiving the most retweets in a given week, with $N = 300$ in the setup of \cite{falkenberg2022growing}. The method is computed as follows:

\begin{enumerate}
    \item Normalize the adjacency matrix: $\mathbf{P} = \mathbf{W} / m$.
    \item Compute the out-degree and in-degree vectors: $\mathbf{r} = \mathbf{P} \mathbf{1}$ and $\mathbf{c} = \mathbf{1}^\top \mathbf{P}$, and their diagonal matrices $\mathbf{D}_r = \mathrm{diag}(\mathbf{r})$, $\mathbf{D}_c = \mathrm{diag}(\mathbf{c})$.
    \item Standardize the matrix using the residuals: 
    \[
    \mathbf{S} = \mathbf{D}_r^{-1/2} (\mathbf{P} - \mathbf{r}\mathbf{c}) \mathbf{D}_c^{-1/2},
    \]
    where $\mathbf{r}\mathbf{c}$ is the outer product of $\mathbf{r}$ and $\mathbf{c}$.
    \item Compute the singular value decomposition (SVD): $\mathbf{S} = \mathbf{U} \Lambda \mathbf{V}^\top$.
    \item Extract the first left singular vector and reweight it as $\mathbf{X} = \mathbf{D}_r^{-1/2} \mathbf{U}_{:,1}$ to obtain an estimate of users' ideological positions.
    \item Rescale $\mathbf{X}$ linearly into the interval $[-1, 1]$.
\end{enumerate}

\begin{figure}[ht!]
	\begin{subfigure}[b]{0.49\textwidth}	
	\includegraphics[width=\textwidth]{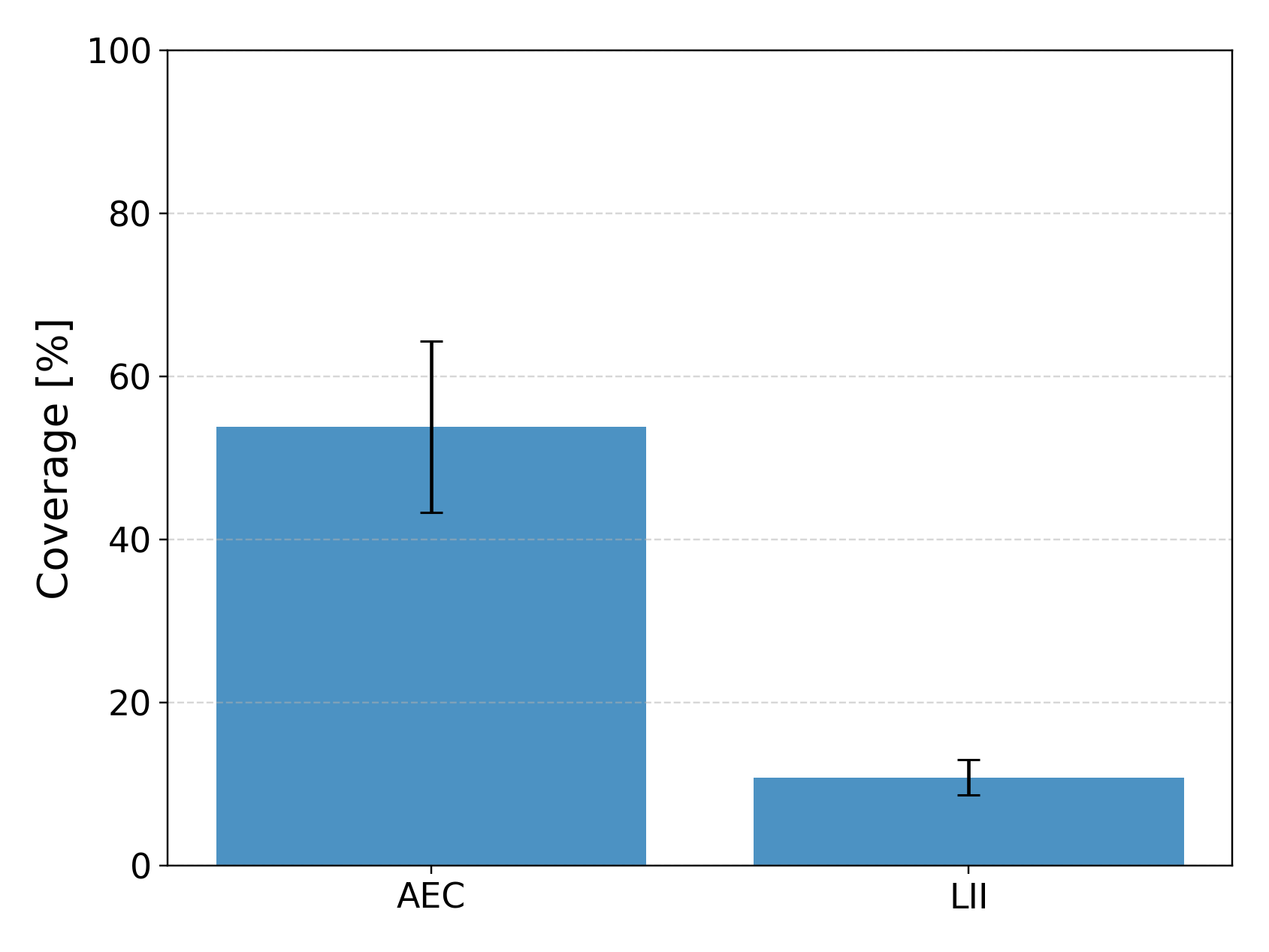}
	\caption{ Coverage }	
	\label{fig:cs_vs_li_coverage}
	\end{subfigure}
	\centering
	\begin{subfigure}[b]{0.5\textwidth}
	\includegraphics[width=\textwidth]{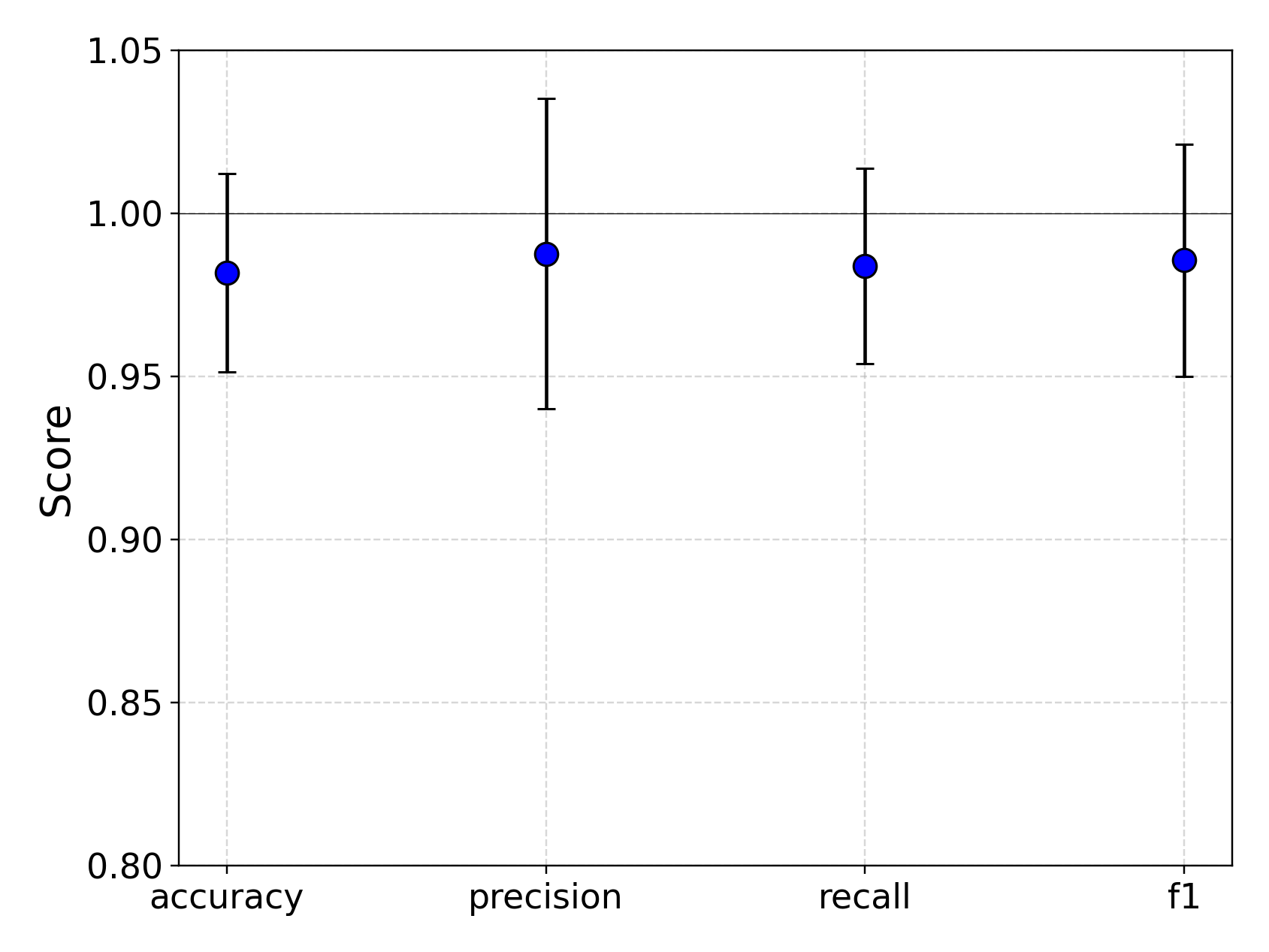}
	\caption{ Performance }	
	\label{fig:cs_vs_li_performance}
	\end{subfigure}
	\caption{\textbf{Comparison between Augmented Echo Chamber (AEC) and Latent Ideology Inference (LII) classification.} Left: Population coverage across weeks, showing that AEC classifies a significantly larger share of users than LII. Right: Classification performance (accuracy, precision, recall, F1) of AEC against LII on the subset of users classified by both methods. Error bars represent standard deviation across weekly networks.}
	\label{fig:cs_vs_li}
\end{figure}

The residual matrix $\mathbf{P} - \mathbf{rc}$ captures deviations from what would be expected under independence between users and elites, that is, if connections were formed purely at random given their marginal interaction probabilities. Standardizing by $\mathbf{D}_r^{-1/2}$ and $\mathbf{D}_c^{-1/2}$ removes scale effects due to user and elite activity levels, allowing the SVD to capture meaningful latent structure. The first singular dimension identifies the principal axis of variation in interaction patterns, which is interpreted as ideological alignment.

We compute LII for every week in our dataset and classify users as climate believers or skeptics based on the thresholding rule from \cite{falkenberg2022growing}.  Figure~\ref{fig:cs_vs_li_performance} shows that AEC achieves a very similar classification to LII when classifying the overlapping set of users, represented by consistently high accuracy, precision, recall, and F1 scores across weeks (error bars represent standard deviation). Despite this similarity in classification quality, Figure~\ref{fig:cs_vs_li_coverage} highlights a key advantage of AEC: it covers a much larger share of the population--nearly four times more on average--while maintaining scalability. This is because AEC leverages chamber structure, allowing it to classify users beyond those who directly interact with elite accounts, making it more effective for mapping broader ideological landscapes.

\section{Configuration model}\label{app:configuration}

To assess the significance of our observations, we compare the observed overlap distribution with that of a random network generated from a null model. We choose the directed configuration model [\cite{hofstad_2016}, Chapter 7] for such a benchmark.  It consists of all possible graphs that share the same \emph{in-} and \emph{out-} degree sequence as our original network. It can be thought as the same network but where all directed edges have been reshuffled in a way that all degrees are preserved.  Since belonging to an audience or a chamber depends only in the positivity of $W_{ij}$, we can work directly with the unweighted adjacency matrix,
\begin{align}
    A_{ij}^* & = \left\lbrace \begin{array}{cc} 1 & \textrm{if }W_{ij}>0\\ 0 & \textrm{if }W_{ij} = 0\end{array}\right. ,   \\
    \kin_i & = \sum_{j} A_{ji}^*, \label{eq:deg-seq-in}\\
    \kout_i & = \sum_{j} A_{ij}^*, \label{eq:deg-seq-out}
\end{align}
where we are denoting by $\bA^*$ the unweighted adjacency matrix of built by our data, \eqref{eq:def-Wij}. We denote by $\bA$ a random adjacency matrix sampled from the configuration model. Mathematically, the probability mass function can be written with Kronecker deltas in the following way, 
\begin{align}
\label{eq:configuration-model}
    p(\bA|\bk^{\textrm{in}}, \bk^{\textrm{out}}) =  \frac{1}{\mathcal{Z}}\prod_{i=1}^N \delta_{\kin_i, \sum_{j}A_{ji}} \delta_{\kout_i, \sum_j A_{ij}}.
\end{align}
Samples of this distribution were obtained by reshuffling the edges of our original network. This choice of randomness results in networks with a locally tree-like structure. More realistic choices can be proposed, but they are harder to control when trying to match with real networks, \cite{Aguirre_L_pez_2021,foster2010communities}. 

Calculating the average value of relevant observables of \eqref{eq:configuration-model} is hard to do analytically, so we choose to work with the related \emph{soft} configuration model for theoretical considerations. In this model, all edges are statistically independent, but with probabilities that match the given degree sequences on average \eqref{eq:deg-seq-in}, \eqref{eq:deg-seq-out},
\begin{align}
\label{eq:soft-configuration-model}
    p_{\textrm{soft}}(\bA|\bk^{\textrm{in}}, \bk^{\textrm{out}}) = \prod_{i\not = j} p(A_{ij}) = \prod_{i \not = j} \p{p_{ij}}^{A_{ij}} (1 - p_{ij})^{1 - A_{ij}}.
\end{align}
where $p(A_{ij} = 1) = p_{ij}$ since $A_{ij} \in \{ 0,1\}$. In order to satisfy the degree constraints, these values must satisfy the following equations,
\begin{equation}
    \begin{aligned}
        \kin_i & = \sum_{j=1}^N p_{ji} \\
        \kout_i & = \sum_{j=1}^N p_{ij}
        \label{eq:degree-constraints}
    \end{aligned}
\end{equation}
This is a set of $2 N$ equations for $N^2 - N$ variables, $p_{ij}$, leading to under-parameterized. We choose these values such that they also maximize the Shannon entropy of the full distribution \eqref{eq:soft-configuration-model}, as it is typical in random graph modelling \cite{coolen2017generating}. This amounts to reducing the problem to $2N$ variables, $\beta^{\textrm{out}}_{i},\beta^{\textrm{in}}_{i} \ge 0$ in such a way that each is associated to one of the constraints. The idea of choosing the maximum entropy distribution is that it is the most unbiased choice for any other observables, \cite{jaynes1957information}. The probability of an edge is then defined in the following way,
\begin{equation}
    \begin{aligned}
        p_{ij} = \frac{\beta^{\textrm{out}}_{i}\beta^{\textrm{in}}_{j}}{1 + \beta^{\textrm{out}}_{i}\beta^{\textrm{in}}_{j}}.
        \label{eq:edge-probability}
    \end{aligned}
\end{equation}
With this choice, \eqref{eq:edge-probability} and \eqref{eq:degree-constraints} constitute a closed set of equations.

Calculating exactly the expected overlap is not trivial since by definition, \eqref{eq:jaccard}, it would be the average of the \emph{ratio} of two random quantities,
\begin{align}
    \expected{q_{ij}^\chamber} = \expected{\frac{|\chamber_i\cap \chamber_j|}{| \chamber_i \cup \chamber_j |}} = \expected{\frac{\sum_\ell \indicator{\ell \in \chamber_i \vee \ell \in \chamber_j}}{\sum_{\ell}\indicator{\ell \in \chamber_i \wedge \ell \in \chamber_j}}},
\end{align}
where we have introduce the indicator function of a statement, $\indicator{ \textrm{statement}}$, defined as $1$ if the statement is true and $0$ otherwise.
We then approximate the expected overlap as the average of the numerator over the average of the denominator of \eqref{eq:jaccard}, 
\begin{equation}
\begin{aligned}
    \expected{q_{ij}^\chamber} &\approx 
    \frac{\expected{\sum_\ell \indicator{\ell \in \chamber_i \vee \ell \in \chamber_j}}}{\expected{\sum_{\ell}\indicator{\ell \in \chamber_i \wedge \ell \in \chamber_j}}} = \\
    & = \frac{\sum_\ell \Prob \s{ \ell \in \chamber_i}\Prob \s{ \ell \in \chamber_j} }{\sum_\ell \Prob \s{ \ell \in \chamber_i} + \Prob \s{ \ell \in \chamber_j} - \Prob\s{\ell \in \chamber_i}\Prob\s{\ell \in \chamber_j}},
\end{aligned}
\end{equation}
where the second identity is true due to the fact that the probability for a given node $\ell$ to be in the chamber of $i$, $\Prob \s{ \ell \in \chamber_i}$, is independent of it being in the chamber of $j$, $\Prob \s{ \ell \in \chamber_j}$, by the independence of edges in \eqref{eq:configuration-model}. It is all a matter then of calculating $\Prob \s{ \ell \in \chamber_i}$, which can also be done thanks to the independence property.
\begin{equation}
\begin{aligned}
    \Prob \s{ \ell \in \chamber_i} &=  1 - \Prob \s{ \ell \not \in \chamber_i}=1 - \Prob\s{\sum_{j = 1}^N A_{ji}A_{ j\ell } = 0}  =\\
    &= 1 - \prod_{j=1}^N\Prob[A_{ji}A_{j\ell} = 0] = 1 - \prod_{j=1}^N \p{1 - \Prob[A_{ji} = 1 \wedge A_{j\ell} = 1]}= \\
    &= 1 - \prod_{j=1}^N \p{1 -  p_{ji} p_{j\ell}} ,
    \label{eq:prob_chamber}
\end{aligned}
\end{equation}

We can do a similar analysis for the overlap similarity between audiences:
\begin{equation}
\begin{aligned}
    \expected{q^\audience_{ij}} & = \expected{\frac{|\audience_i\cap \audience_j|}{| \audience_i \cup \audience_j |}} = \expected{ \frac{\sum_\ell \indicator{\ell \in \audience_i \vee \ell \in \audience_j}}{\sum_\ell\indicator{ \ell \in \audience_i \wedge \ell \in \audience_j}}} \\
    & \approx
    \frac{\sum_\ell \Prob\s{\ell \in \audience_i}\Prob\s{\ell \in \audience_j}}{\sum_\ell \Prob\s{\ell \in \audience_i} + \Prob\s{\ell \in \audience_j} - \Prob\s{\ell \in \audience_i}\Prob\s{\ell \in \audience_j}} ,
\end{aligned}
\end{equation}
where here everything depends on the probability of a node being part of a given audience, which is exactly the probability of the edge.
\begin{align}
    \Prob\s{\ell \in \audience_i} = \Prob\s{A_{\ell  i } = 1} = p_{\ell i}
    \label{eq:prob_audience}
\end{align}
While in theory by solving for the values of the $\beta^{\textrm{in}}_i,\beta^{\textrm{out}}_i$'s from \eqref{eq:edge-probability} is enough, it is still not an explicit solution in terms of the degree distributions. For the case where the distributions are not broad enough, that is $k_\textrm{max}\ll N$, one can make the following approximation,
\begin{equation}
    \begin{aligned}
        \beta^{\textrm{in}}_i & = \frac{\kin_i}{\sqrt{N \expected{k}}}, \\
        \beta^{\textrm{out}}_i & = \frac{\kout_i}{\sqrt{N \expected{k}}},
    \end{aligned}
\end{equation}
which leads to 
\begin{equation}
    \begin{aligned}
        \Prob\s{\ell \in \audience_i} & = \frac{\kout_\ell \kin_i}{N\expected{k}} + \mathcal{O}\hspace{-1mm}\left(\frac{1}{N^2}\right),\\
        \Prob\s{\ell \in \chamber_i} & = \frac{\kin_i \kin_\ell}{N} \frac{\expected{(\kout)^2}}{{\expected{k}}^2} + \mathcal{O}\hspace{-1mm}\left(\frac{1}{N^2}\right).
    \end{aligned}
    \label{eq:approximations}
\end{equation}
where $N \expected{k} = \sum_i \kin_i = \sum_i \kout_i$ and $N \expected{(\kout)^2} = \sum_i (\kout_i)^2$. Since probabilities cannot be larger than $1$, these expressions show their limit of validity. For distributions with a rapidly decaying tail, like an exponential one, these constitute a very good approximation. While they may not be valid for broader tails like power laws, the qualitative picture remains the same.  With these approximations, one can derive the following expressions for the average overlaps,
\begin{align}
    \expected{q_{ij}^\chamber} &\approx \frac{\kin_i \kin_j }{N(\kin_i + \kin_j)} \frac{\expected{(\kout)^2}}{ {\expected{k}}^2} \frac{ \expected{(\kin)^2} } {\expected{k}} \\
    \expected{q_{ij}^\audience} & \approx \frac{\kin_i \kin_j }{N(\kin_i + \kin_j)} \frac{ \expected{(\kout)^2}}{ {\expected{k}}^2} ,
\end{align}
so then if  we have $\expected{(\kin)^2}>\expected{k}$, then it follows that
\begin{align}
    \expected{q_{ij}^\chamber}> \expected{q_{ij}^\audience}.
\end{align}
In most social networks, the degree variance, $\expected{(\kin )^2}$, is much larger than the average degree, $\expected{k}$. Therefore, we expect the chamber overlap to be significantly larger than the audience overlap. Notice also the out-degree distribution plays a significant role and a large variance in the out-degree (as observed) leads to typically larger values of chambers and overlaps. 

\section{Unsupervised clustering}\label{app:unsupervised_communities}
We take a spectral clustering approach to obtain the communities of the leading users. 
We consider the similarity Laplacian matrix, $\mathbf{L} = \mathbf{D} - \mathbf{Q}$, where $\mathbf{Q}$ is the overlap similarity matrix aggregated over all the weeks in the dataset, and $\mathbf{D}$ is a diagonal matrix with the degree sequence of $\bm{Q}$ in the diagonal, i.e., we have that $\mathbf{D}_{ii} = \sum_{j} q_{ij}$ and $D_{ij} = 0$ for $i \neq j$.

By the spectral properties of $\bm{L}$, we know that connected networks have the smallest eigenvalue of $0$ followed by positive eigenvalues. The second smallest eigenvalue $\lambda_2 > 0$, also called the \textit{algebraic connectivity}, reflects how well-connected the network is. Its associated eigenvector, $\bm{u}_2$, also called the \textit{Fiedler vector}, is typically used to partition networks into two non-trivial communities that minimize that cut size, i.e., it minimizes the sum of weights between the groups \cite{mohar1997some}. 

\begin{figure}[t!]
    \centering
	\includegraphics[width=0.99\textwidth]{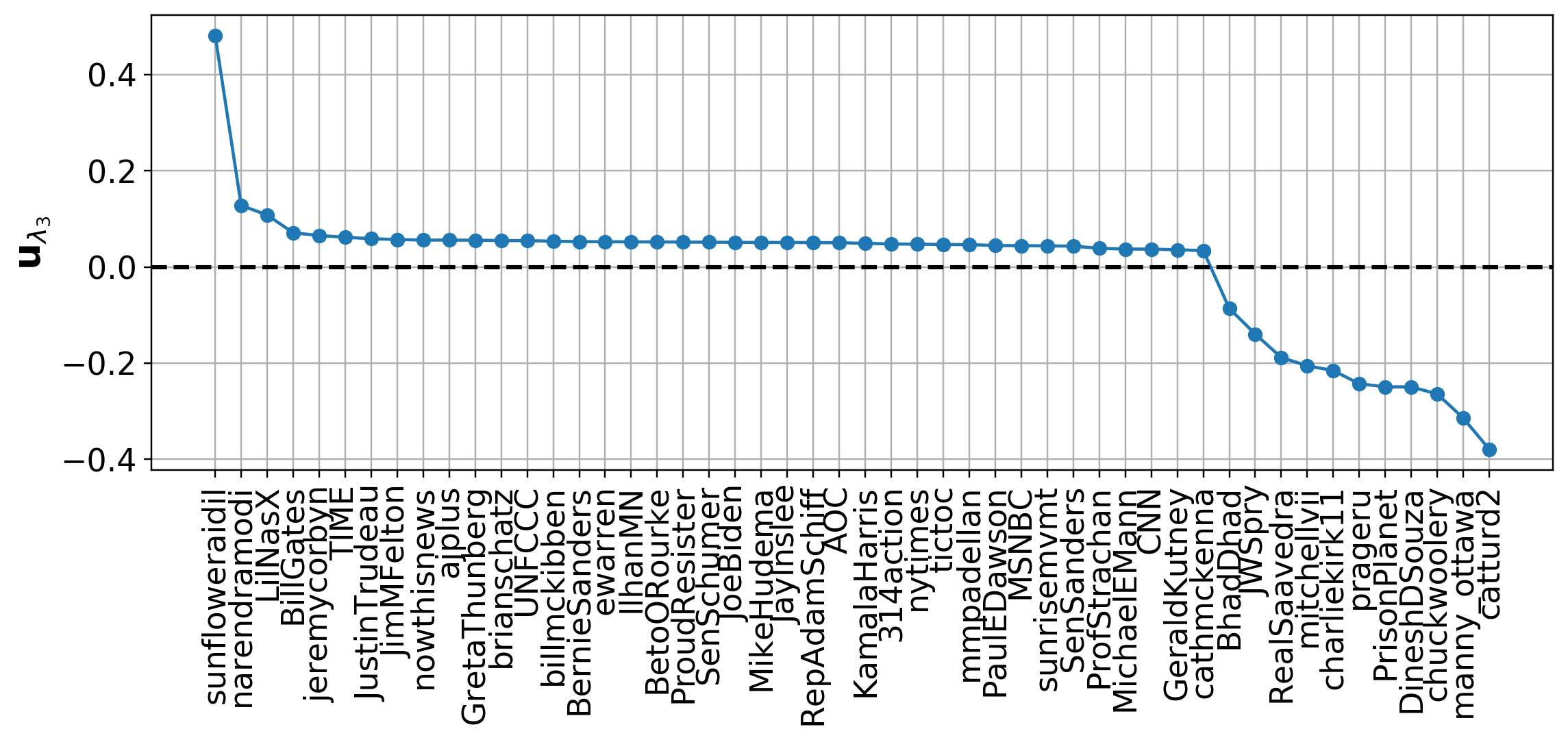}
	\caption{\textbf{Eigenvector associated with $\lambda_3$, the third leading eigenvalue, sorted by magnitude.} We identify between climate believers and skeptics with the sign of their corresponding component in $\mathbf{u}_{\lambda_3}$. }
	\label{fig:eigenvector}
\end{figure}

In the case of the Laplacian of the aggregate overlap similarity matrix, the second smallest eigenvector separates the satellite users (namely, \textit{BhadDhad}, \textit{narendramodi}, \textit{sunfloweraidi} and \textit{LilNasX}) from the rest of the network. While the separation associated with the second eigenvalue separates the network into almost independent connected components, it does not capture the groups suggested by the bimodality of the overlap distribution. Instead, we partition the leading users into two groups by taking the eigenvector associated with the \textit{third} smallest eigenvalue, $\lambda_3$. We identify the entries of $\bm{u}_3$ that are \textit{greater than zero} as \textit{climate believers} while the entries \textit{smaller than zero} as \textit{climate skeptics}. In Fig. \ref{fig:eigenvector}, we show the individual entries of $\bm{u}_3$, where each entry corresponds to a leading user.

\section{Comparing chambers and audiences \label{app:comparison}}

Many recent studies about polarization on social media have used a first-neighbors approach to characterize the structure of the interaction networks \cite{williams2015network,barbera2015tweeting, cinelli2021echo, jiang2021social, chen2021polarization}. In X (formerly Twitter), retweets have been used extensively as a proxy for endorsement \cite{barbera2015birds, gaumont2018reconstruction}, so retweet networks have been used to study the assortative structure of several conversations. In this work's terminology, these studies have studied the \textit{audience} of a set of users to quantify polarization, echo chambers, and other nontrivial social structures. We argue that the chamber is, under certain conditions, a more robust tool to quantify such social structures, both conceptually and mathematically. The difference is that looking at the chamber corresponds to looking at the information sources of the audience, a second-neighbors approach.

\begin{figure}[ht!]
	\begin{subfigure}[b]{0.49\textwidth}	
	\includegraphics[width=\textwidth]{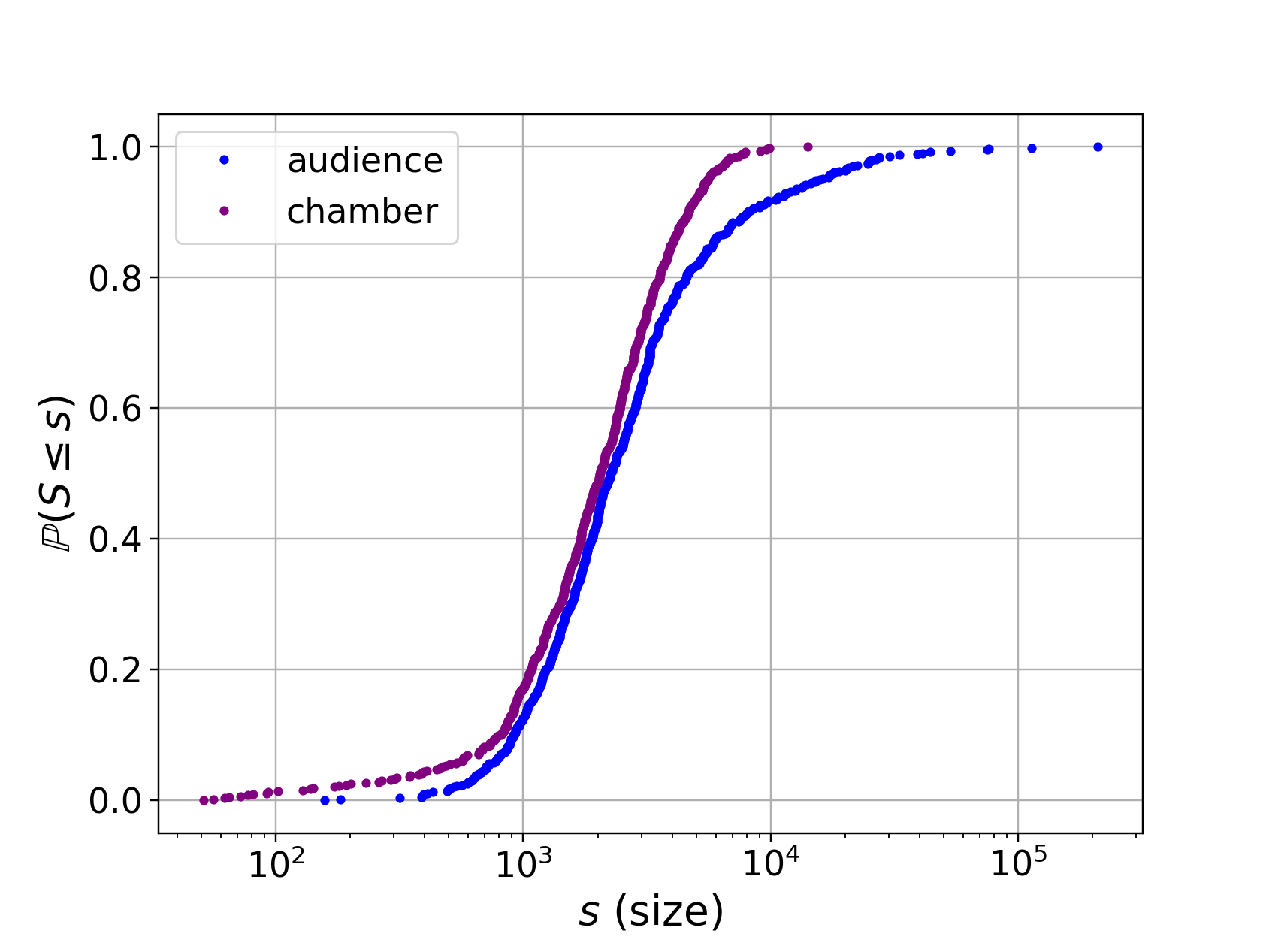}
	\caption{ Cumulative distribution function }	
	\label{fig:size_cdf}
	\end{subfigure}
	\centering
	\begin{subfigure}[b]{0.49\textwidth}
	\includegraphics[width=\textwidth]{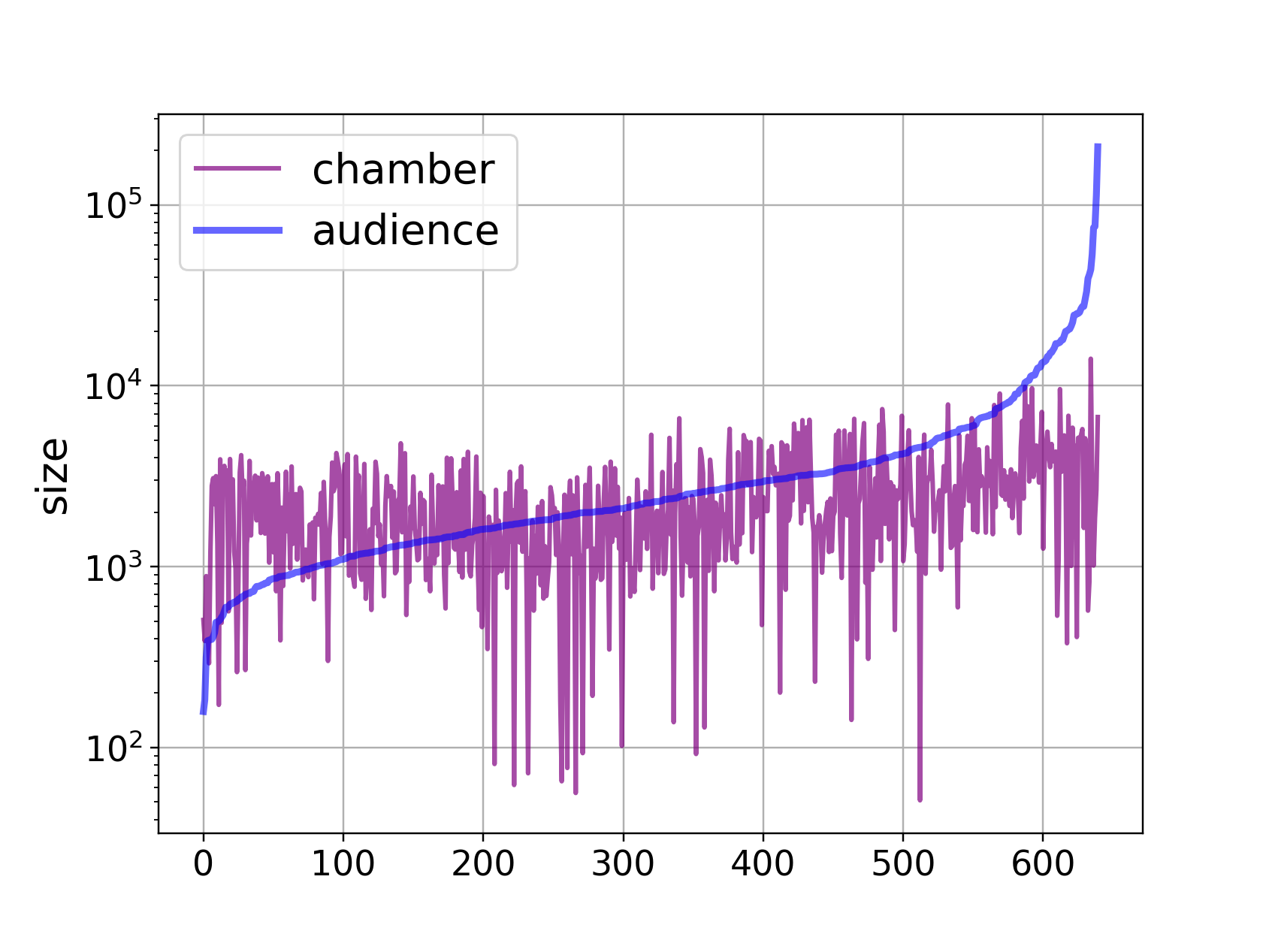}
	\caption{ Audience-ordered sizes  }	
	\label{fig:size_correlations}
	\end{subfigure}
	\caption{\textbf{Chamber and audience sizes.} \textit{a)} Cumulative distribution function. We construct each distribution by concatenating the sizes of the chambers (purple) (audiences (blue)) of the leading users, $\topusers^\Delta(t)$, for every week $t$ on the dataset. The audience size distribution ($s = 4668 \pm 11472$) has a significantly longer right-tail than the chambers ($s = 2425 \pm 1682$). \textit{b)} Audience-sized ordered chamber (purple) and audience (blue) sizes from smallest to largest for the leading users for all weeks. We find a small correlation ( $\rho = 0.26$) between the audience and chamber sizes.}
	\label{fig:size_distributions}
\end{figure}

From a mathematical perspective, for the null model, the \textit{expected overlap} between the \textit{audiences} of two high-impact users is significantly lower than the expected overlap between chambers (see Appendix \ref{app:configuration} for details). This suggests that the chamber overlap should typically have a higher signal-to-noise ratio. Moreover, in the case of the climate change retweet network, the distribution of chamber sizes has a significantly lower spread than that of the audience, as we show in Fig. \ref{fig:size_cdf}. In other words, many chambers are roughly the same size, while the audience sizes vary greatly. The overlap similarity is informative whenever the size of the two sets compared are of the same order, so using the chamber gives us more reliable results than the audience. For two arbitrary sets $A$ and $B$ where $|A| \ll |B|$, the Jaccard overlap $ |A \cap B |/|A \cup B| \approx |A|/|B| $, which indicates the relative size of $A$ with respect to $B$ and not their overlap.

Besides mathematical arguments, we argue that the audience and the chamber are fundamentally different objects. By construction, one would expect the chamber to be tightly coupled to the audience because the audience determines the chamber. For instance, if a user has a big audience, we expect her to also have a big chamber. Surprisingly, the correlation between audience and chamber sizes is very low ($\rho = 0.26$), suggesting that they are loosely coupled (see Fig. \ref{fig:size_correlations}).

\begin{figure}[ht!]
    \centering
    \includegraphics[width=0.7\textwidth]{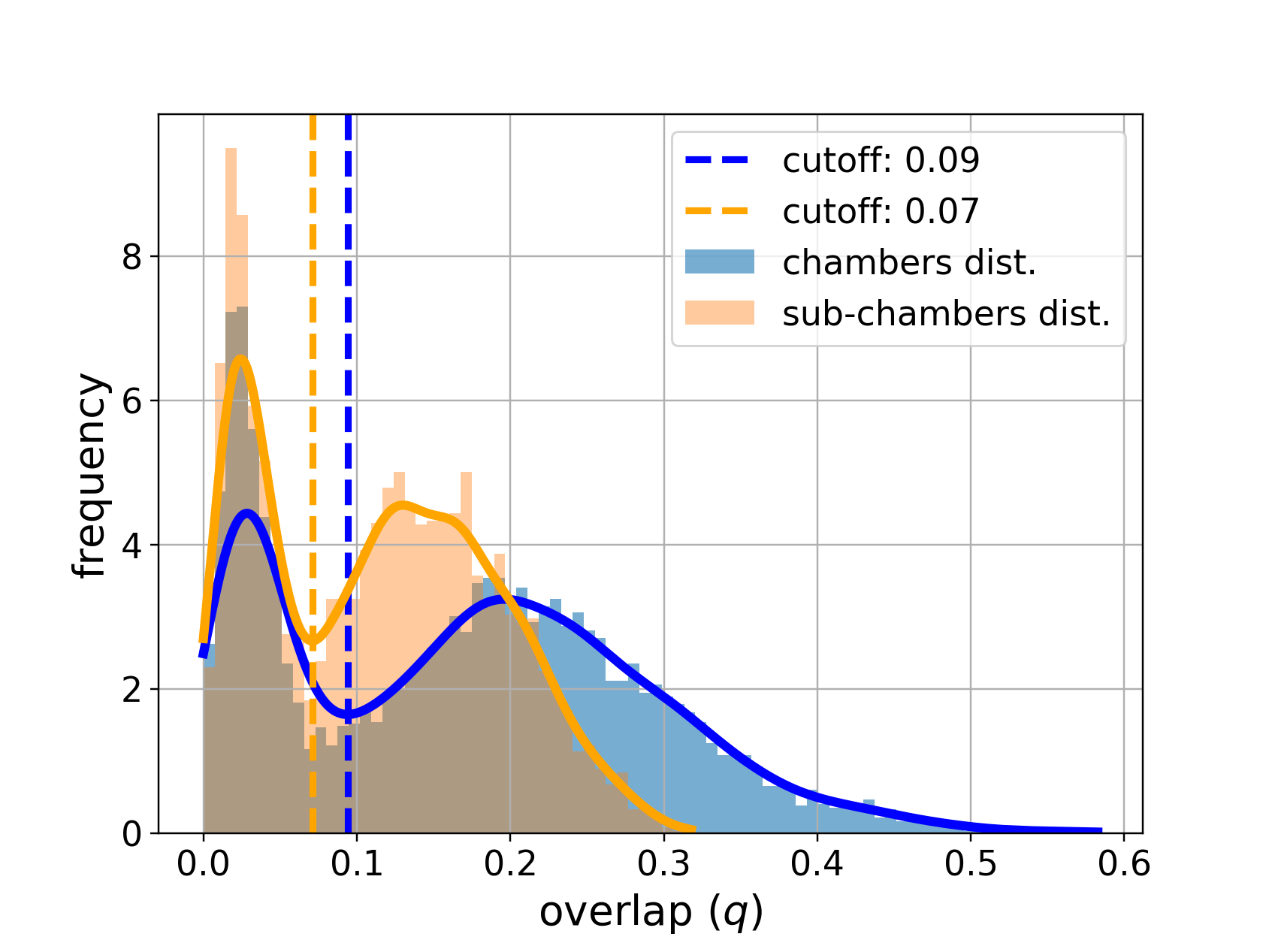}
    \caption{\textbf{Aggregate chamber and subchamber overlap distributions.} We construct the aggregate chamber (blue) (subchamber (orange)) distribution by concatenating the overlap pairs, $q_{ij}^t$, for every week $t$ on the dataset -- i.e., $\bm{q} = (q_{ij}^t)_{t, i<j}$. A subchamber differs from a chamber in that we remove the overlapping users of the audience before constructing between the subchamber (see main text for details on their construction). While the first peak of the subchamber distribution $q_{off}^s = 0.03 \pm 0.02$ is almost identical to that of the chamber distribution ($q_{off} = 0.04 \pm 0.02$), the second peak, $q_{in} = 0.16 \pm 0.5$ is lower and has a smaller spread ($q_{in} = 0.23 \pm 0.08$). However, both peaks are clearly bimodal and have similar cutoff values.} 
    \label{fig:subchamber_dist}
\end{figure}

We further analyze the chambers' behavior by removing their audiences' coupling. When comparing high-impact users $i$ and $j$, we remove the common audience members of $i$ and $j$ and construct their chambers \textit{without} them. In Fig. \ref{fig:subchamber_dist}, we compare the overlap similarity distributions with and without removing the common audience members. Both are bimodal. We observe that the expected overlap of the second peak decreases but is still significantly high. Moreover, both distributions share the same bimodal structure with a similar cutoff, indicating that we only removed redundant information by removing the coupling caused by the audiences. This suggests that the chamber is more robust to missing information.

Finally, we argue that the audience is a collection of information \textit{consumers}, whereas a chamber is a collection of information \textit{sources}. On the one hand, every audience member consumes information from the leading user in a traditional one-to-many fashion. On the other hand, the audience consumes information from the chamber in a many-to-many fashion, where the ideological coherence comes from considering the chamber as a whole and not its members. Thus, the overlap between chambers reflects the similarity between the audiences' many-to-many information channels, giving us a proxy of the ideological (dis)similarities between leading users/leading users.

\section{Leading users description}\label{app:leading users_description}

We show in Table \ref{tab:leading users_description} the list of leading users and their main characteristics. 

\begin{table}[ht]
\centering
\resizebox{\textwidth}{!}{\begin{tabular}{|l|rrrrll|}
\toprule
leading user &  Persistence &  Total impact &  Median impact &  Chamber size & Ideology (spectral) & Ideology (manual) \\
\midrule
BernieSanders  &           24 &        410442 &           5093 &                   35811 &             believers &           believers \\
LilNasX        &           10 &        320133 &          10301 &                    6583 &             believers &               other \\
PaulEDawson    &           39 &        282040 &           7349 &                   88025 &             believers &           believers \\
AOC            &           23 &        243209 &           7481 &                   40302 &             believers &           believers \\
CNN            &           30 &        158068 &           3077 &                   38677 &             believers &               other \\
ewarren        &           23 &        116018 &           3519 &                   29140 &             believers &           believers \\
MikeHudema     &           29 &        112690 &           3345 &                   49874 &             believers &           believers \\
sunfloweraidil &            7 &        102276 &           3951 &                    1989 &             believers &           believers \\
SenSanders     &           19 &         94747 &           2936 &                   26297 &             believers &           believers \\
RepAdamSchiff  &            8 &         92173 &           9855 &                   21765 &             believers &           believers \\
ajplus         &           21 &         78273 &           3179 &                   20646 &             believers &               other \\
sunrisemvmt    &            8 &         77320 &           1802 &                   11992 &             believers &           believers \\
GeraldKutney   &           31 &         73966 &           2180 &                   51325 &             believers &           believers \\
KamalaHarris   &           12 &         69900 &           4365 &                   16269 &             believers &           believers \\
IlhanMN        &            7 &         69250 &           4806 &                   13716 &             believers &           believers \\
JimMFelton     &            9 &         66397 &           5102 &                   12995 &             believers &           believers \\
GretaThunberg  &            9 &         64787 &           5480 &                   20537 &             believers &           believers \\
BetoORourke    &           12 &         59042 &           2455 &                   16247 &             believers &           believers \\
nowthisnews    &           17 &         58796 &           3094 &                   18299 &             believers &               other \\
nytimes        &           15 &         49493 &           2174 &                   19417 &             believers &               other \\
JayInslee      &           16 &         44600 &           2594 &                   20705 &             believers &           believers \\
narendramodi   &            7 &         43984 &           4203 &                    3669 &             believers &           believers \\
JoeBiden       &            9 &         40170 &           3084 &                    8333 &             believers &           believers \\
ProudResister  &            9 &         38873 &           3048 &                   12921 &             believers &           believers \\
brianschatz    &           10 &         36951 &           2959 &                   10740 &             believers &           believers \\
mmpadellan     &            7 &         34335 &           4047 &                   13277 &             believers &           believers \\
jeremycorbyn   &            7 &         33573 &           2569 &                   11454 &             believers &           believers \\
UNFCCC         &           16 &         32000 &           2047 &                   24574 &             believers &           believers \\
SenSchumer     &            9 &         31598 &           3098 &                   10875 &             believers &           believers \\
JustinTrudeau  &            8 &         30146 &           3582 &                   10281 &             believers &           believers \\
ProfStrachan   &           14 &         28860 &           1786 &                   26876 &             believers &           believers \\
billmckibben   &            8 &         26331 &           2398 &                   16595 &             believers &           believers \\
314action      &           11 &         25086 &           2215 &                    8853 &             believers &           believers \\
tictoc         &            7 &         22500 &           2588 &                   10495 &             believers &               other \\
cathmckenna    &           10 &         22239 &           1762 &                   15957 &             believers &           believers \\
BillGates      &            8 &         19071 &           2400 &                    4872 &             believers &           believers \\
MichaelEMann   &            9 &         17947 &           1966 &                   18596 &             believers &           believers \\
TIME           &            7 &         16815 &           2039 &                    7665 &             believers &               other \\
MSNBC          &            7 &         14651 &           1882 &                    8401 &             believers &               other \\
PrisonPlanet   &           15 &        154928 &           5180 &                   29046 &              skeptics &            skeptics \\
DineshDSouza   &           13 &         75806 &           4515 &                   14267 &              skeptics &            skeptics \\
charliekirk11  &            7 &         65267 &           8427 &                   16081 &              skeptics &            skeptics \\
BhadDhad       &            7 &         60552 &           2111 &                    2352 &              skeptics &               other \\
chuckwoolery   &            9 &         55050 &           3691 &                   13988 &              skeptics &            skeptics \\
RealSaavedra   &            9 &         53977 &           3109 &                   11506 &              skeptics &            skeptics \\
catturd2       &            7 &         53477 &           3783 &                   14776 &              skeptics &            skeptics \\
prageru        &            8 &         32300 &           2825 &                   14277 &              skeptics &            skeptics \\
manny\_ottawa   &           12 &         28981 &           2256 &                    8308 &              skeptics &            skeptics \\
JWSpry         &           13 &         21588 &           1690 &                   16439 &              skeptics &            skeptics \\
mitchellvii    &            8 &         17876 &           1914 &                    5769 &              skeptics &            skeptics \\
\bottomrule
\end{tabular}}
\caption{Description of the $M = 50$ leading users defined in Eq. (\ref{eq:persistent_leading_users}) in terms of their persistence, $\Delta_i$, their cumulative impact, $\sum_t w_i^t$, their median impact, $Med(w_i^t)_t$, their aggregate chamber size, $| \cup_t \chamber_i^t |$, and their ideology based on manual labeling and the spectral clustering algorithm described in Appendix \ref{app:unsupervised_communities}. According to the spectral clustering algorithm, users are ordered by ideology and from largest to smallest impacts.}
\label{tab:leading users_description}
\end{table}

\section{Descriptions and definitions for polarization and echo chambers}\label{app:echochamber_desc}

In the main text, we argue that while many authors have studied polarization and echo chambers in online social networks, there is no consensus on these concepts. Intuitively, the polarization between two groups refers to the tendency of each group to interact mostly with its members, rejecting the interaction with the other group. Similarly, an echo chamber promotes in-group communication while rejecting out-group communication. Moreover, members in an echo chamber tend to systematically reinforce their beliefs because the information posted by a given member of the echo chamber \textit{bounces} back to that member through the flow of communication with other users within it.

In Tables \ref{tab:polarization_desc} and \ref{tab:echochamber_desc}, we describe different notions of polarization and echo chambers, respectively, based on the literature on online social networks.

\begin{table}[ht]
    \centering
    \begin{tabularx}{\textwidth}{|l|X|X|c|}
    \toprule
    \makecell{\textbf{Concept} \\ \textbf{(Polarization)}}     &  \textbf{Description} & \textbf{Equation}  \\
    \midrule
    \makecell{Polarization \\ index \cite{becatti2019extracting}}  
    &  
    ``A high value indicates that a user systematically interacts with people belonging to her/his same coalition rather than with users from different alliances, but we disregard any information concerning the content of the shared news, and we do not take into account how it is amplified by being shared within the same group of people.''
    & 
    \begin{equation*}
        \arg\max_\alpha \frac{\lvert \mathcal{N}_i \cap C_\alpha \rvert}{\lvert \mathcal{N}_i \rvert},
    \end{equation*}
    $\mathcal{N}_i$: set of leading users $i$ interacted with 
    
    $\mathcal{C}_\alpha$ the set of leading users in community $\alpha$ \\ \hline
    \makecell{Adaptive E-I \\ index \cite{chen2021polarization}}
    &
    ``We measure polarization as the relative density of in-group agreement to out-group agreement. We assess these patterns on our endorsement networks, which are a subtype of general communication networks where all ties are publicly conveyed indications of agreement''
    &
    \begin{equation*}
        \frac{n_{\alpha\alpha}^t + n_{\beta\beta}^t - ( n_{\alpha\beta}^t + n_{\beta\alpha}^t )}{n_{\alpha\alpha}^t + n_{\beta\beta}^t + ( n_{\alpha\beta}^t + n_{\beta\alpha}^t )},
    \end{equation*}
    the notation is the same as in Eq. (\ref{eq:polarisation}). \\ \hline
    \makecell{Random walker \\ controversy measure \\ (RWC) \cite{garimella2018quantifying}}
    &
    ``Controversial and polarized topics induce graphs with clustered structure, representing different opinions and points of view''
    
    ``The (random walker) measure captures the intuition of how likely a random user on either side is to be exposed to authoritative content from the opposing side.''
    &
    \begin{equation*}
        p_{\alpha \alpha}p_{\beta \beta} - p_{\alpha\beta}p_{\beta\alpha},
    \end{equation*}
    $p_{\alpha\beta}$: probability that a random walker starting in community $\alpha$ ends in community beta. \\
    \bottomrule
    \end{tabularx}
    \caption{Description of the polarization concept used as in the literature, as well as their associated mathematical equation.}
    \label{tab:polarization_desc}
\end{table}

\begin{table}[ht]
    \centering
    \begin{tabularx}{\textwidth}{|l|X|}
    \toprule
    \textbf{Concept} &  \textbf{Description}  \\
    \midrule
    Echo chamber \cite{bruns2017echo} & 
    ``An echo chamber forms when a group of participants chooses to preferentially connect, to the exclusion of outsiders. The more fully formed this [social] network is, the more isolated from the introduction of outside views is the group, while the views of its members can circulate widely within it'' \\ \hline
    Echo chamber \cite{garimella2018quantifying} & 
    ``Opinions or beliefs stay inside communities created by like-minded people, who reinforce and endorse the opinions of each other'' \\ \hline
    Echo chamber \cite{bakshy2015exposure} & 
    ``Individuals are exposed only to information from like-minded individuals'' \\ \hline 
    Echo chamber \cite{cinelli2021echo} & 
    ``Environments in which users' opinions, political leaning, or beliefs about a topic are reinforced due to repeated interactions with peers or sources with similar tendencies and attitudes. Selective exposure and confirmation bias may explain the emergence of echo chambers on social media.'' \\ \hline
    Echo chamber \cite{sunstein1999law} & 
    ``One of the consequences of the group polarization phenomenon is to cast new light on an old point, to the effect that social homogeneity can be quite damaging to good deliberation. When people hear echoes of their voices, the consequence may be far more than support and reinforcement. Another consequence is that particular forms of homogeneity can be breeding grounds for unjustified extremism, even fanaticism'' \\
    \bottomrule
    \end{tabularx}
    \caption{Description of the main notions of an echo chamber in the literature. In general, echo chambers form as a \textit{consequence} on group polarization. \textit{Because} group polarization exists, members inside those groups are only exposed to information of like-minded individuals and \textit{hear echoes} of their voices.}
    \label{tab:echochamber_desc}
\end{table}

\end{document}